\documentclass[preprint,12pt]{elsarticle}

\usepackage{graphicx}
\usepackage{amssymb}
\usepackage{gensymb}
\usepackage{textgreek}
\usepackage[dvipsnames]{xcolor}
\usepackage{fontenc}
\usepackage[acronym]{glossaries}
\usepackage{siunitx}
\usepackage{url}
\usepackage{subfig}
\usepackage{multirow}
\usepackage{xcolor}

\newacronym{cots}{COTS}{Commercial Off-The-Shelf}
\newacronym{aasi}{AaSI}{Aalto Spectral Imager}
\newacronym{radmon}{RADMON}{Radiation Monitor}
\newacronym{epb}{EPB}{Electrostatic Plasma Brake}
\newacronym{fmi}{FMI}{Finnish  Meteorological  Institute}
\newacronym{eps}{EPS}{Electrical Power Subsystem}
\newacronym{tt&c}{TT\&C}{Telemetry, Telecommand \& Communication}
\newacronym{adcs}{ADCS}{Attitude Determination \& Control Subsystem}
\newacronym{obdh}{OBDH}{Onboard Data Handling}
\newacronym{obc}{OBC}{On Board Computer}
\newacronym{gui}{GUI}{Graphical User Interface}
\newacronym{mcc}{MCC}{Mission Control Center}
\newacronym{eqm}{EQM}{Engineering-Qualification Model}
\newacronym{fm}{FM}{Flight Model}
\newacronym{vtt}{VTT}{VTT Technical Research Centre of Finland}
\newacronym{fpi}{FPI}{Fabry--P\'{e}rot Interferometer}
\newacronym{pcb}{PCB}{Printed Circuit Board}
\newacronym{uhf}{UHF}{Ultra High Frequency}
\newacronym{gps}{GPS}{Global Positioning System}
\newacronym{epscb}{EPSCB}{Electrical Power System Control Board}
\newacronym{fmea}{FMEA}{Failure mode and effects analysis}
\newacronym{sdr}{SDR}{Software Defined Radio}
\newacronym{tle}{TLE}{Two Line Element}
\newacronym{sdram}{SDRAM}{Synchronous Dynamic Random Access Memory}
\newacronym{i2c}{I$^2$C}{Inter Integrated Communication}
\newacronym{uart}{UART}{Universal Asynchronous Receiver Transmitter}
\newacronym{spi}{SPI}{Serial Peripheral Interface}
\newacronym{pate}{PATE}{Particle Telescope}
\newacronym{leo}{LEO}{Low Earth Orbit}
\newacronym{rf}{RF}{Radio Frequency}
\newacronym{sme}{SME}{Small \& Medium Enterprises}
\newacronym{bcr}{BCR}{Battery Charge Regulators}
\newacronym{pcm}{PCM}{Power Conditioning Module}
\newacronym{radef}{RADEF}{RADiation Effects Facility}
\newacronym{ads}{ADS}{Antenna Deployment System}

\newcommand{\jaan}{\color{black}}

\journal{Acta Astronautica}

\begin{document}

\begin{frontmatter}

\title{Aalto-1, multi-payload CubeSat: design, integration and launch}
\author[1]{J.~Praks}
\author[1]{M.~Rizwan~Mughal
\footnote{M. Rizwan Mughal is also associated with Electrical Engineering Department, Institute of Space Technology, Islamabad, Pakistan, Correspondence: rizwan920@gmail.com}}
\author[2]{R.~Vainio}
\author[3]{P.~Janhunen}
\author[3]{J.~Envall}
\author[2]{P.~Oleynik }
\author[4]{A.~N\"asil\"a}
\author[1]{H.~Leppinen}
\author[1]{P.~Niemel\"a}
\author[1,5]{A.~Slavinskis}
\author[2]{J.~Gieseler}
\author[3]{P.~Toivanen}
\author[1]{T.~Tikka}
\author[1]{T.~Peltola}
\author[1]{A.~Bosser}
\author[1]{G.~Schwarzkopf}
\author[1]{N.~Jovanovic}
\author[1]{B.~Riwanto}
\author[3]{A.~Kestil\"a}
\author[2]{A.~Punkkinen}
\author[6]{R.~Punkkinen}
\author[6]{H.-P.~Hedman}
\author[6]{T.~S\"antti}
\author[7]{J.-O.~Lill}
\author[8]{J.M.K.~Slotte}
\author[9]{H.~Kettunen}
\author[9]{A.~Virtanen}

\address[1]{Department of Electronics and Nanoengineering, Aalto University School of Electrical Engineering, 02150 Espoo, Finland}
\address[2]{Department of Physics and Astronomy, 20014 University of Turku, Finland}
\address[3]{Finnish Meteorological Institute, Space and Earth Observation Centre, Helsinki, Finland}
\address[4]{VTT Technical Research Centre of Finland Ltd, Espoo, Finland}
\address[5]{Tartu Observatory, University of Tartu, Observatooriumi 1, 61602 T{\~o}ravere, Estonia}
\address[6]{Department of Future Technologies, 20014 University of Turku, Finland}
\address[7]{Accelerator Laboratory, Turku PET Centre, \AA{}bo Akademi University, 20500 Turku, Finland}
\address[8]{Physics, Faculty of Science and Technology, \AA{bo} Akademi University, 20500 Turku, Finland}
\address[9]{Department of Physics, P.O.Box 35, 40014 University of Jyvaskyla, Finland}

\begin{abstract}

The design, integration, testing and launch of the first Finnish satellite Aalto-1 is briefly presented in this paper. Aalto-1, a three-unit CubeSat, launched into Sun-synchronous polar orbit at an altitude of approximately 500 km, is operational since June 2017. It carries three experimental payloads: \gls{aasi}, \gls{radmon} and \gls{epb}.
\gls{aasi} is a hyperspectral imager in visible and near-infrared (NIR) wavelength bands, \gls{radmon} is an energetic particle detector and \gls{epb} is a de-orbiting technology demonstration payload. The platform was designed to accommodate multiple payloads while ensuring sufficient data, power, radio, mechanical and electrical interfaces. The design strategy of platform and payload subsystems consists of in-house development and commercial subsystems. The CubeSat Assembly, Integration \& Test (AIT) followed Flatsat$-$\gls{eqm}$-$\gls{fm} model philosophy for qualification and acceptance.
 
The paper briefly describes the design approach of platform and payload subsystems, their integration and test campaigns and spacecraft launch. The paper also describes the ground segment \& services that were developed by Aalto-1 team.

\end{abstract}


\end{frontmatter}

\section{Introduction}
\label{S:1}

Nowadays, there is an increased interest towards small satellite missions due to advances in \gls{cots} technology miniaturization. Traditionally, the classification of  small satellites is only based on their mass but the CubeSat standard also takes into consideration the volume~\cite{CalPoly2009}. Over the past decade, the applications of small satellites in general and CubeSats in particular have increased manifold due to the availability of low-cost design, testing and launch possibilities~\cite{Frischauf2018,Peters2015,Tkatchova2018,Salt2013}. Initially perceived for training and educational activities, the applications of CubeSats have expanded in vast application areas in the past few years \cite{survey_nanosatellite}. Example application areas include remote sensing, Earth observation, disaster management, science, astronomy, space weather and technology demonstration etc. \cite{doi:10.2514/6.1998-5255, RefWorks:28,RefWorks:18,RefWorks:37,Selva2012,Santilli2018,SEYEDABADI}.

The abundant availability of \gls{cots} components with faster development cycles has led to the NewSpace movement \cite{newspace}. This approach has led the transformation of CubeSat missions from educational and technology demonstration to real missions with potentially risky but higher commercial and science return \cite{sweeting, RefWorks:27, Poghosyan2017}.
A large number of commercial applications using CubeSats have evolved in the past few years with a promising future scope of commercial applications \cite{Frischauf2018,Peters2015,Tkatchova2018,Salt2013}. Until now, more than one thousand CubeSats have been launched into space \cite{thousand_CubeSat}. However, the forecast suggests an exponential increase in nanosatellite launches every year \cite{Crusan2019}. 

There has also been great advancement in the technology development for nano and microsatellites \cite{9183926}. A number of innovative platforms have been designed and demonstrated in space \cite{ali2012tile,7015716,mughal_intra,Ali2018}. Due to technology miniaturization, the capability of CubeSat platforms has been ever increasing \cite{mughal2014plug, ali2014innovative}. The current CubeSat missions are able to demonstrate innovative platforms with high power generation, precise attitude pointing and higher data downlink capabilities with potential to compete with their bigger satellite counterparts. 

During the past decade, the worldwide trend of the first satellite by each university or \gls{sme} has been designing and launching relatively less complex single-unit (1U) CubeSat for capability demonstration. In contrast, we at Aalto university followed a more challenging approach, i.e., designing a multi-payload CubeSat with student teams. The mission objective was to build and launch a spacecraft with focus on science, imaging and de-orbiting technology demonstration while also providing hands-on educational training. This paper presents detailed design aspects of the Aalto-1 CubeSat with a capability description of payloads and the platform to accomplish the mission objectives. The in-orbit demonstration and lessons learned are presented in an accompanying paper~\cite{a1_inorbit_2020}. 

This paper is organized as follows:  Section~\ref{sec:mission_objective} briefly introduces mission objectives and requirements, Section~\ref{mission_and_project_implementation} presents the mission design, project implementation and educational outcomes, section~\ref{space_segment} presents space segment design and implementation,  section~\ref{sec:paylaods} presents all payloads: their specifications and designs, Section~\ref{sec:platform} introduces design approach of the platform subsystems, Section~\ref{sec:integration} presents the integration \& testing, section~ \ref{sec:ground}  focuses on ground segment and Section~\ref{sec:discussion} concludes the paper.

\section{Mission objectives}
\label{sec:mission_objective}
{\jaan The Aalto-1 satellite project was initiated from Aalto University student’s aspiration to make the first satellite mission in Finland. The idea was supported by teachers and developed during a special assignment in Space Technology course in 2010 spring semester in the form of feasibility study of the satellite. 
The goal of the course was set to develop a realistic satellite concept which should be possible to implement (at least partly) by students. It was required that the main payload should be developed in Finland and it should be connected to Aalto University curriculum. During the feasibility study this was translated to the goal to build first Finnish satellite with Earth Observation payload. }

{\jaan For the university}, the main driver for the Aalto-1 project was to provide hands-on education in space engineering, science and entrepreneurship, while taking advantage of the NewSpace movement \cite{newspace,Frischauf2018,Peters2015,Tkatchova2018,Salt2013} {\jaan and harness the enthusiasm of building the first national satellite}. It was envisioned, that in addition to satellite development, students will also learn to work with experienced space scientists and develop connections to industrial partners.{\jaan 
The mission was largely financed and led by Aalto University and integrated to Aalto space technology curriculum. 

Despite a main goal of building, launching and operating first national satellite, the proposed payload selection introduced complex technology demonstration and science goals.}  The first feasibility study built the satellite concept around four payload candidates and established a consortium for building a 3U Cubesat. The study also derived the main mission requirements. As an outcome of the feasibility study, the satellite mission and platform was to be developed by Aalto University students and  payloads were to be contributed by partner organisations. 

The main payload candidate was a spectral Earth Observation imager, \gls{aasi}, based on technology developed by \gls{vtt}. This further led to wider spectral device offering for space applications by \gls{vtt}. Another payload candidate, a radiation monitoring device, later called \gls{radmon}, was proposed by a team from University of Turku and University of Helsinki. The third payload candidate, selected by the study, was e-sail experiment device, \gls{epb}, which was already in development at \gls{fmi} for ESTCube-1 CubeSat mission \cite{estcube1_lessons, Aalto1_Technical}.
In the original feasibility study, a vibration monitoring system was also proposed. However, the idea was later abandoned as impractical for a rather monolithic nanosatellite.

Neither of the selected payloads had flight heritage. Moreover, \gls{aasi} and \gls{epb} main technology was never demonstrated in space before for proposed purpose. Earth Observation with tunable \gls{fpi} was a novel concept and also deorbiting a satellite with electrostatic force by using a tether was never attempted before. This provided a technical challenge and scientific novelty for the project. The project consortium, which consisted of Aalto University, University of Turku, University of Helsinki, VTT and Finnish Meteorological Institute, decided to build a multi-payload mission. In a retrospective, one can say that this decision elongated the project significantly and enforced also several compromises in the design due to contradicting requirements. It took slightly over five years from the first idea to \gls{fm} completion. The overarching Aalto-1 mission objective was to build a satellite to carry out in-orbit demonstration of \gls{aasi}, \gls{radmon} and \gls{epb} experiments, each of them with specific mission objectives. 

\gls{aasi}'s main objective was to demonstrate the operation of a tunable \gls{fpi}-based spectral imager for Earth Observation in the space environment. {\jaan The \gls{fpi} technology developed at \gls{vtt} allowed to build for the first time freely tunable spectral EO camera to nanosatellite form factor.} As a minimum, the instrument was required to take wavelength calibration measurements and record a spectrum of at least six wavelengths of a cloud-free land. For a full demonstration, the instrument was required to take measurements to investigate wavelength stability, thermal effects and long-term degradation of filters, optics, the sensor and other components along with demonstrating various operation modes. 

\gls{radmon}'s main objective was to operate and calibrate a CubeSat-compatible radiation detector which registers protons in nine energy channels with threshold energies of 10 -- 40~MeV and electrons in five energy channels with threshold energies of 1.5 -- 12 MeV.

\gls{epb}'s main objective was to deploy a tether and then charge it to estimate the force exerted by the Coulomb drag between the tether's electric field and the Earth's ionosphere, as well as to demonstrate de-orbiting by keeping the tether charged for an extended period of time. This novel propulsion concept was (and currently still is) never demonstrated in space. By now two launched nanosatellites, ESTCube-1 and Aalto-1 have made attempts to deploy this system. However, in the near future AuroraSat-1, Foresail-1  and ESTCube-2 are heading towards similar goals using the same technology \cite{Iakubivskyi2019}.

Detailed mission requirements kept developing along the project and were not  documented in detail. Therefore, it can be said, that the mission was technology driven as it often happens in CubeSat missions. However, the proper feasibility study in the very beginning and well established consortium helped to keep the focus on results. 

The finally launched satellite followed closely the original plan of payloads and functionality, but seriously underestimated requirements due to many constraints including time and resources.

\section{Mission design and implementation}
\label{mission_and_project_implementation}

In order to satisfy the payload in-orbit demonstration requirements, Aalto-1 was required to be launched to a polar orbit with an altitude of at least 500~km. A polar orbit provides sufficient conditions to estimate the Coulomb drag force~\cite{estcube1_adcs_high_2014} and allows for \gls{radmon} to measure at various latitudes, including the South Atlantic Anomaly. Polar orbit also allows coverage in Finland and provides good opportunities for Earth Observation. 

The attitude requirements were set by all payloads, but dominated by \gls{epb} requirements.The lower limit of altitude was required by \gls{epb}. In lower altitudes, the atmospheric drag might dominate the de-orbiting impact which makes electrostatic drag estimation difficult. The highest altitude limit was set by 25-year orbital decay requirement for space debris mitigation. The \gls{epb} experiment requires spinning satellite of hundreds of degrees per second in order to provide centrifugal force for tether deployment~\cite{eas_A1}. The angular momentum was to be provided in steps: spin up the satellite, deploy the tether, spin up again, etc. \gls{aasi} requires nadir pointing during image acquisition and \gls{radmon} requires attitude knowledge, but the requirement was not critical. 
Another notable requirement for the mission was surface conductivity requirement by \gls{epb} to keep spacecraft potential during Coloumb drag experiment. 
The satellite was designed for two years in the orbit, which was estimated as sufficient time to carry out all experiments. 
The mission design in terms of energy and thermal budget was flexible, as it was decided that payload duty cycles can be adjusted in-orbit according to the need.  

The satellite operation from Aalto University was one of the key mission requirements. For this purpose, the ground segment was developed. The ground station includes UHV, VHF and S-band steerable antennas and associated transceivers. The mission operation software was designed and implemented by Aalto students.

\begin{figure}[h!]
    \centering
    \includegraphics[width=\columnwidth]{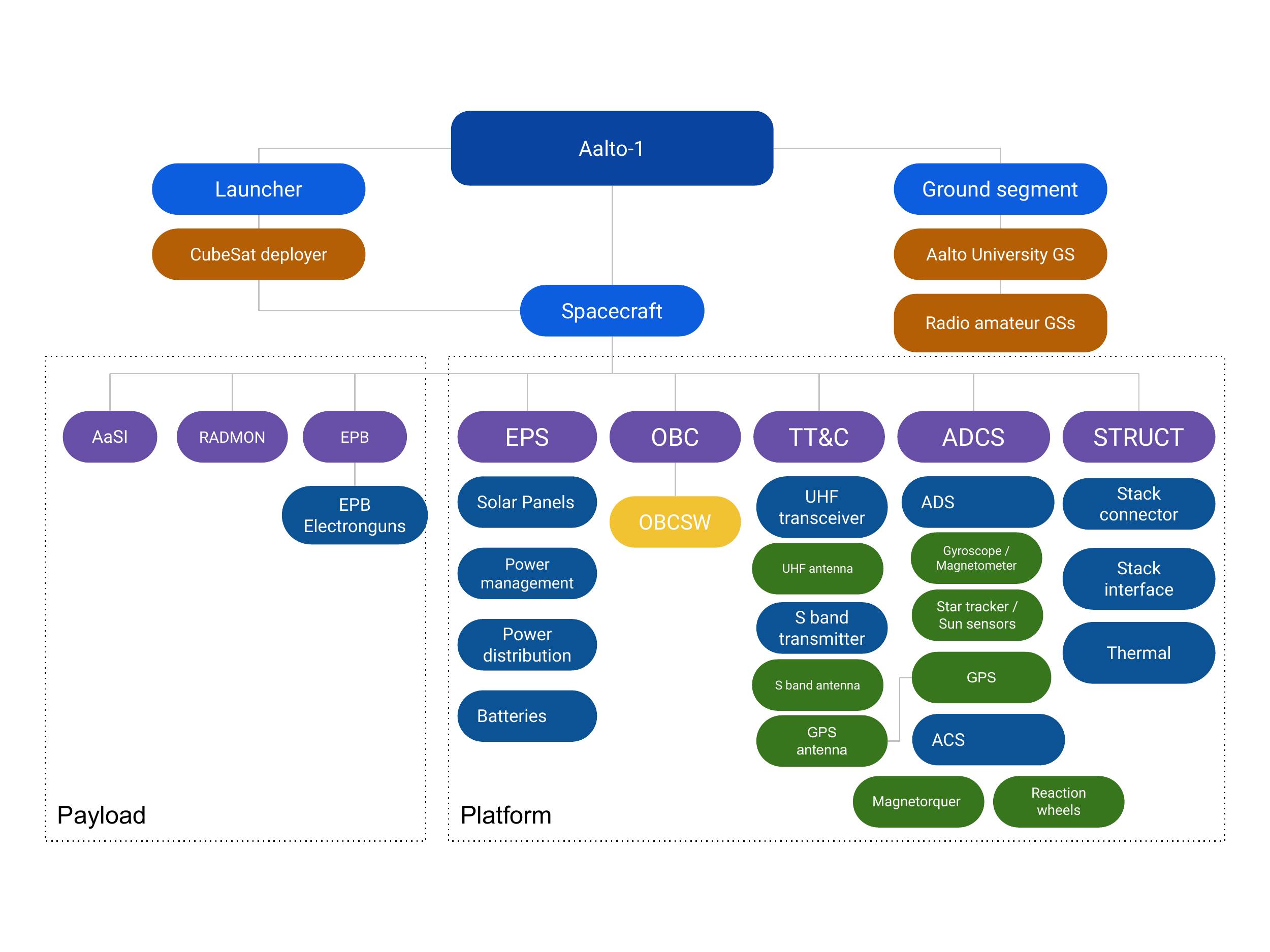}
    \caption{Aalto-1 product tree}
    \label{fig:product_tree}
\end{figure}

The product tree of Aalto-1 mission with ground segment, space segment and launch segment description is shown in Fig~\ref{fig:product_tree}. 

\subsection{Project Implementation}

After successful feasibility study in spring 2010, the satellite project was quickly funded and supported by Aalto MIDE (Multidisciplinary Institute of Digitalisation and Energy)\cite{MIDE}. The project was also formally organized by establishing posts for project responsible professor, project coordinator, Steering Group, and Scientific Advisory Board. Under the official project umbrella, student groups developed their own organization for building the satellite and ground segment. Thematic student teams were in most cases oriented on single subsystem development or a single topic. The quality assurance was maintained as a separate independent branch as it is practiced in bigger satellite projects. During the semester, student teams had weekly meetings and decisions were made by team-leader meetings. 

Thanks to the available funding, it was possible to hire few doctoral students, provide summer trainee positions and occasional master thesis positions with salary (usual practice in Finnish Universities). The doctoral students formed a backbone of the project which helped accumulate the salient knowledge. Many subsystems were developed as a master thesis project. 

Payload teams formed separate project structure in their home organizations and their team leaders were part of the Scientific Advisory Board. Satellite bus and payload developments were financially independent and applied for funds independently. Several satellite project students made their master thesis with payload team. 

The project schedule was built to mimic larger space projects, where the main project phases were separated by milestone reviews. The Preliminary Design Review was arranged in November 2011, Critical Design Review in May 2013, Test Readiness Review in May 2015 and Flight Readiness Review in January 2016. Several smaller reviews were arranged along the project. The flight model of the satellite was delivered to Netherlands in May 2016. 

Review panels were assembled from space technology professionals and CubeSat team members from other universities. Both, documentation based review format (in the beginning of the project) and presentation based review format (towards the end of the project) were used. 
 
Flatsat, \gls{eqm} and \gls{fm} model policy with fast iterative development model was implemented in the project. A single Flight Qualification Model (FQM) approach was considered in the beginning of the project, but it proved to be impractical. The students were inexperienced and learned most efficiently by making prototypes and hardware versions, therefore rapid iterations and frequent hardware models proved to be more efficient than waterfall design.
  
The main challenge for the project was to find and keep the knowledge in the team during multi year project. Student teams were volatile and documentation often incomplete, despite a requirement for documentation to retrieve study credits. 
The fact that key persons were hired and committed to the project, helped the project to continue. Constant support by the university and project organization was also highly important. The project was also well aligned with university goals: it was able to produce degrees, research papers and positive publicity. Aalto University procured and financed, along with few sponsoring partners, also the satellite launch, the first in Finland. 

\subsection{Educational outcomes}

The student work was incorporated to student's individual studies mainly via special assignments, bachelor and master thesis projects and also as a part of doctoral studies.
The main challenge was to align project needs and project documentation with teaching and outcome assessment in situation where most of the work was done in groups.

During the project evolved a documentation and reporting approach where project documentation was used for grading and individual contributions were assessed by self evaluation and peer reviews. Assessment was done on the basis of provided snapshot of the evolving documentation and it was required that the documentation was available for entire project. The final grade was assigned by supervising professor \cite{Praks15Aalto1Edu}. 
Far more that 100 students contributed in the design and development of the satellite. However, the contribution varied from single semester participation in meetings to several years of design and implementation.
Around 12 Master level and 28 Bachelor level thesis were conducted in the satellite design and development activity during the coarse of the project.
By now, also three doctoral dissertations are defended based mainly on Aalto-1 satellite related topics \cite{Leppinen18PhD, OsamaPhD,KestilaPhD} and several are still on the way. Additionally more than 10 Master level thesis were conducted at partner institutes related to the design and development of payloads. The outcomes and results were also published in many scientific conferences and journals in the field. The project gathered a lot of media attention which led to the awareness of space technology and small satellites in vast areas \cite{Praks2015}.

Many of the Aalto-1 students became space engineers and scientists at partner institutions. A subgroup of Aalto-1 students established the ICEYE company, which builds operates a fleet of Synthetic-Aperture Radar (SAR) satellites. Another group formed {\jaan Reaktor Space Lab company, which specializes on nanosatellite missions.}

\section{Space Segment design and implementation}
\label{space_segment}
\begin{figure}[h!]
\centering
\includegraphics[width=\columnwidth]{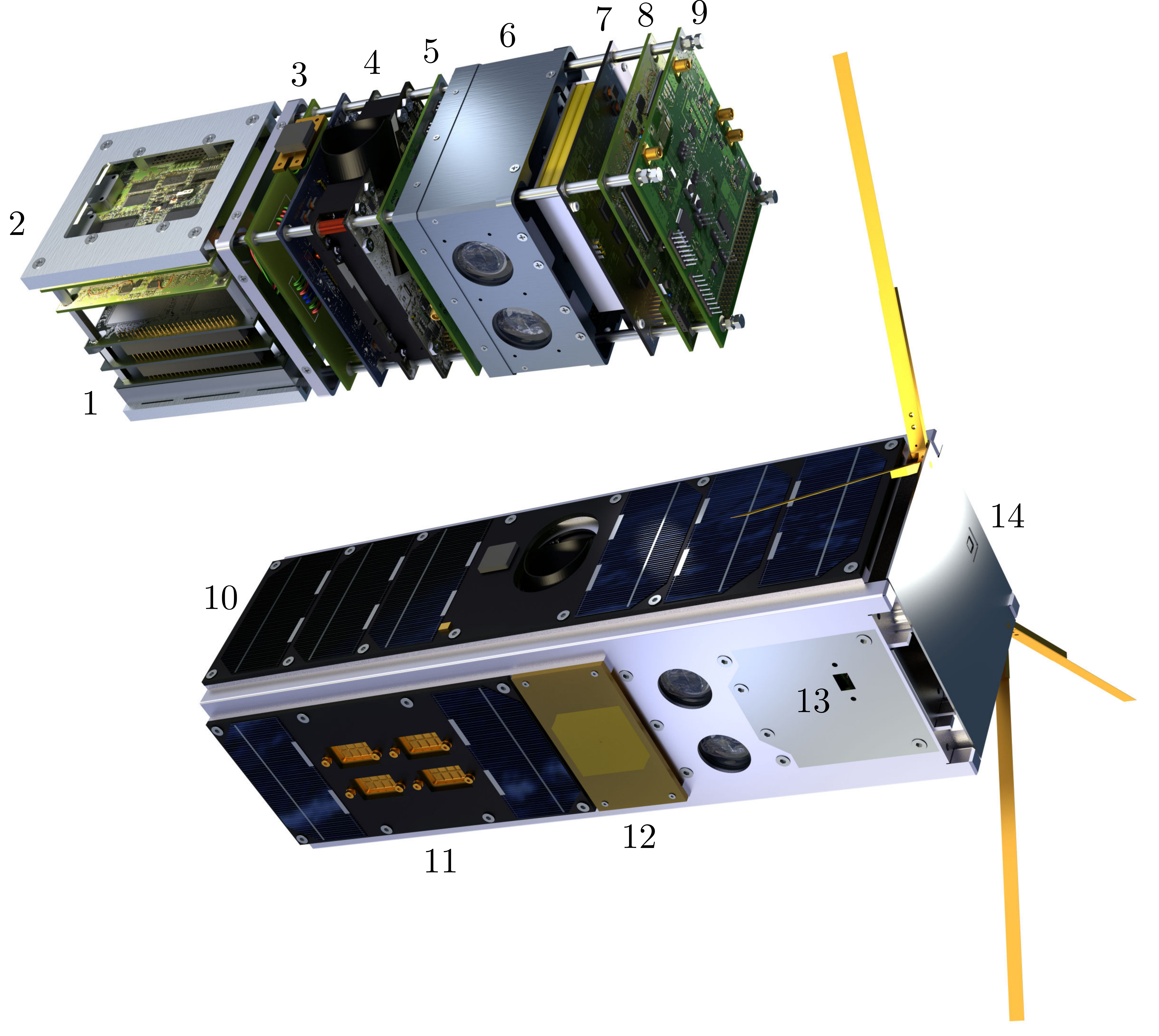}
\caption{The structure and subsystems of Aalto-1 satellite. The highlighted subsystems are: 1) Radiation Monitor (RADMON), 2) Electrostatic Plasma Brake (EPB) 3) Global Positioning System's (GPS's) antenna and stack interface board, 4) Attitude Determination and Control System (ADCS), 5) GPS and S-band radio, 6) Aalto Spectral Imager (AaSI), 7) Electrical Power System (EPS), 8) On-Board Computer (OBC), 9) Ultra High Frequency (UHF) radios, 10) solar panels, 11) electron guns for EPB, 12) S-band antenna, 13) debug connector, and 14) UHF antennas.}
\label{fig:a1internals}
\end{figure}

{\jaan The feasibility study and} preliminary design analysis {\jaan proposed}  3U CubeSat platform to carry out the mission. {\jaan The CubeSat platform was selected because it provided affordable access to space and also available commercial subsystems for inexperienced team. The payloads were designed concurrently with the satellite platform, \gls{aasi} and \gls{radmon} were entirely new designs whereas \gls{epb} development was already started for ESTCube-1 satellite \cite{estcube1_lessons}. 

The 3U satellite platform was designed and manufactured mainly by students of Aalto University. However, in early stage of the project it was decided that electrical power system and attitude system should be procured from commercial provider. The main reason for that was the reliability concern of fresh designs.

The satellite design, as shown in Fig.~\ref{fig:a1internals}, features 3U CubeSat body, 3-axis stabilization, body mounted solar panels, deployable UHF antennas, several cameras and openings for payloads. Electronics of the satellite is accommodated in two electronic stacks, connected by cabling. The Long Stack features all the avionics and \gls{aasi} payload. The Short Stack accommodates \gls{radmon} and \gls{epb}. The main reason for this separation was the design decision to align the \gls{epb} reel motor rotation axis with satellite rotation axis in spinning mode.
}  

The student designed Aalto-1 platform consists of a in-house developed cold-redundant \gls{obc} running Linux, \gls{uhf} and S-band radios, a navigation system based on \gls{gps}, {\jaan aluminium structure, solar panels, Sun sensors, \gls{ads}} and commercially procured\gls{eps} and \gls{adcs}. 
\begin{figure}[h!]
    \centering
    \includegraphics[width=\columnwidth]{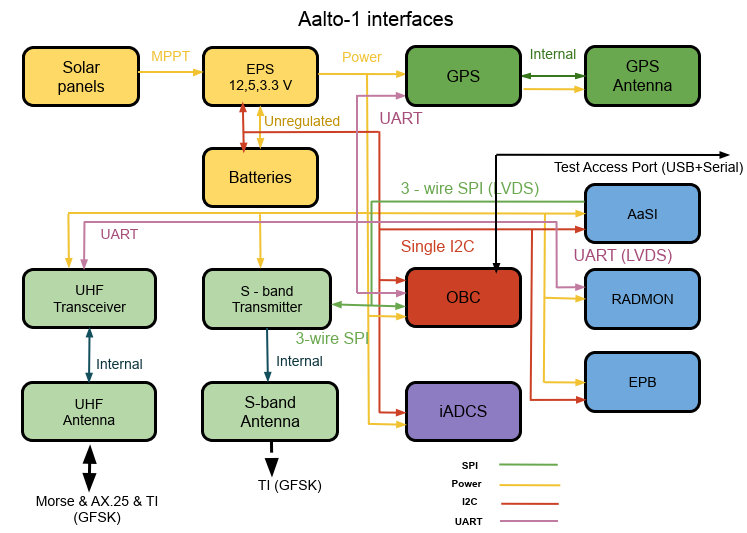}
    \caption{A block diagram of digital, RF and power interfaces}
    \label{fig:obc_interfaces}
\end{figure}

An overview of the Aalto-1 power, data and \gls{rf} interfaces is presented in Fig.~\ref{fig:obc_interfaces}. The power interface provides regulated voltage levels (3.3~V, 5~V and 12~V) to the satellite avionics and payloads. Several digital interfaces including \gls{i2c}, \gls{spi} and \gls{uart} have been implemented which are controlled by the \gls{obc}. 
\newpage
\section{Payloads}
\label{sec:paylaods}
The instrument design technique, mass, volume, electrical and mechanical interfaces and key design challenges of the Aalto-1 payloads is described in detail in this and subsequent sections.
\subsection{Radiation monitor}
\begin{figure}[h!]
\centering
\subfloat[The \gls{radmon} radiation monitor assembly.] {{\hspace*{-1cm}\includegraphics[width=8.5cm]{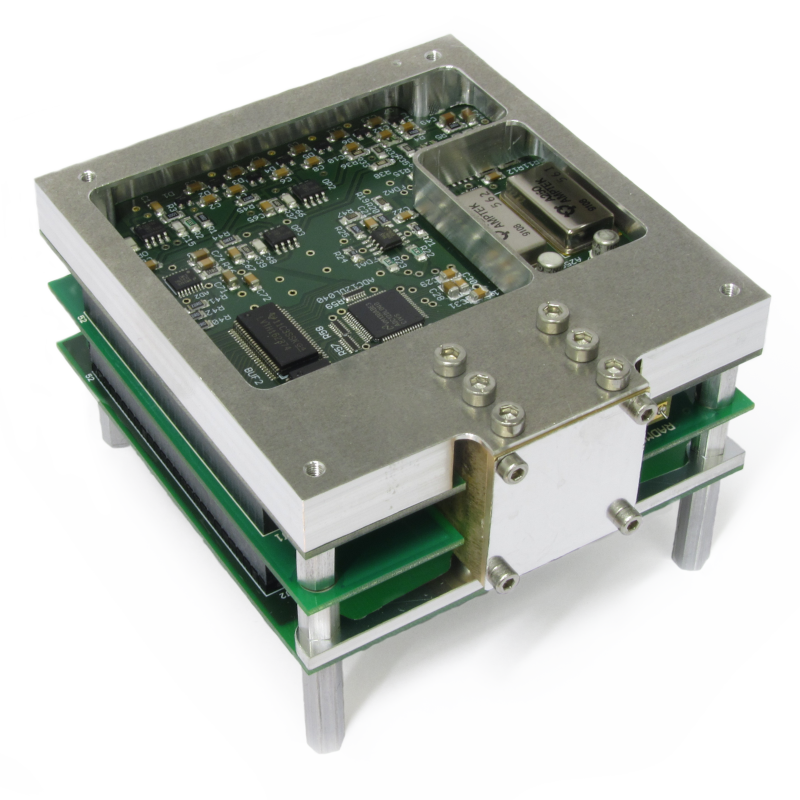}}}%
\qquad
\subfloat[The detector unit.] {{\hspace*{0cm} \includegraphics[width=4cm]{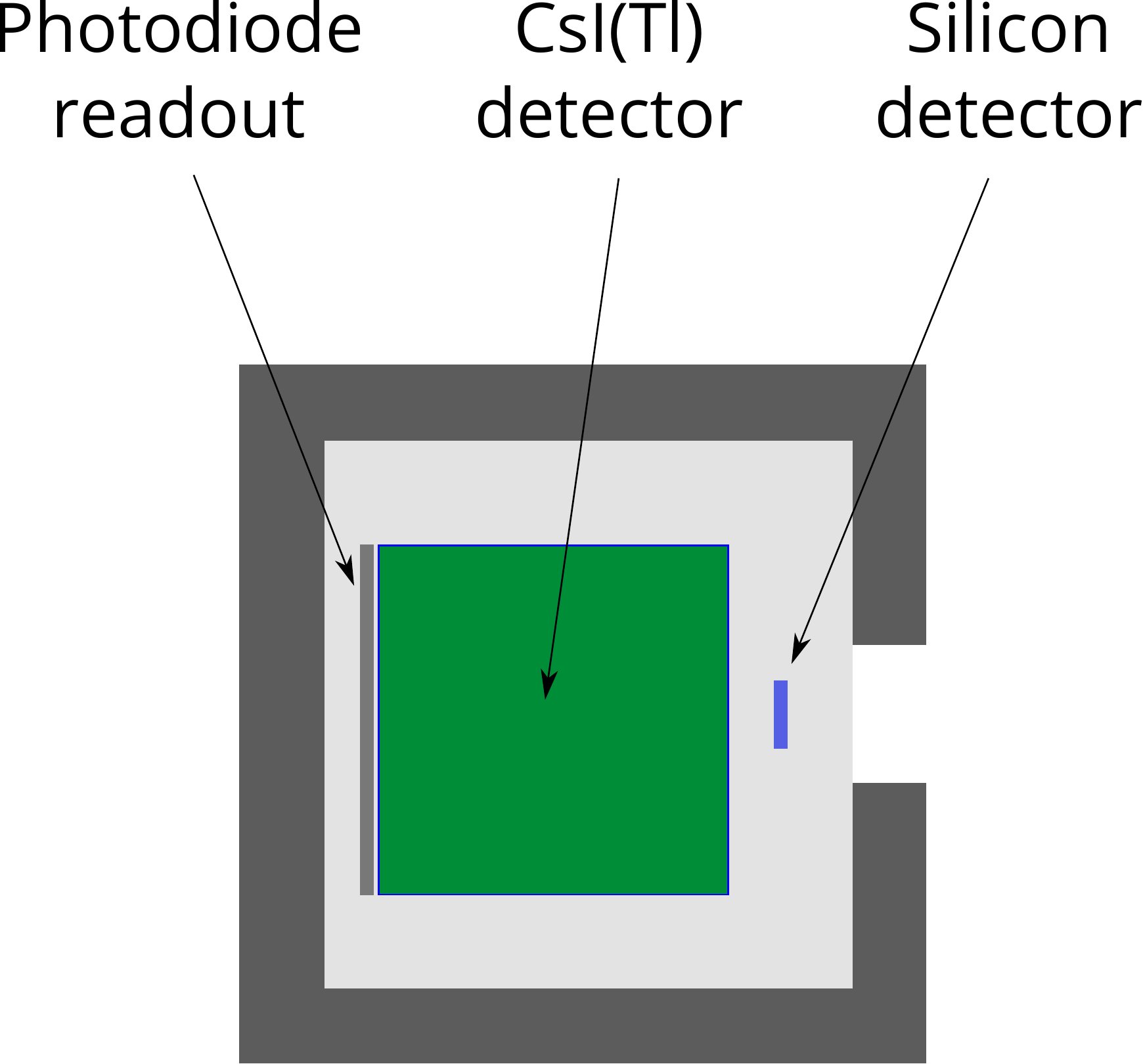}
}}%
\caption{The assembly compounds three printed circuit boards and a brass container with a detector unit inside. An aluminum entrance window in front of the brass container covers the detector unit. A scintillator with a readout photodiode and a silicon detector are housed within the brass container.}
\label{fig:radmon}
\end{figure}

The \gls{radmon} instrument \citep{Peltonen2014} is a compact low-power radiation monitor. It has envelope dimensions of about 4$\times$9$\times$10 $\text{cm}^3$, a mass of 360~g and a power consumption of 920~mW. The spacecraft supplies both +5~V and +12~V to the instrument. The instrument consists of a detector assembly inside a brass casing, a signal processing board, a digital board, and an electrical power board. Three boards are connected by a 52-pin internal bus running through all of the boards (see Fig.~\ref{fig:radmon}(a)). The instrument is integrated in the short stack of the satellite with another bus connector as well as with four spacers placed in the corners of the PCB stack. The bus connector also provides the electrical interface to the satellite.
	
The detector unit consists of a rectangular 2.1$\times$2.1$\times$0.35~$\text{mm}^3$ silicon detector and a 10$\times$10$\times$10~$\text{mm}^3$ CsI(Tl) scintillation detector that are enclosed by the brass casing determining the acceptance aperture (see Fig.~\ref{fig:radmon}(b)). The casing has an aluminum entrance window that protects the detector stack from low-energy charged particles and photons. The scintillator has a thin polytetrafluoroethylene (PTFE) wrapping on five sides and has a readout photodiode on the sixth side. We have used a Hamamatsu S3590-08 PIN silicon photodiode with dimensions of 10$\times$10 $\text{mm}^2$ and a depletion thickness of about 0.3~mm.  The silicon detector has a biased guard ring and a floating one. The passive area of the silicon detector extends to about 0.7~mm around the active spot. Two detectors produce electrical signals for a standard \textDelta{E} -- E analysis aimed at the determination of particle species and the energy deposited in the detector. A coincidence logic prevents the registration of particles coming from outside the aperture and bremsstrahlung X-rays generated in the brass container.

The aluminum window sets thresholds for electron detection at about 1~MeV and for proton detection at about 10~MeV. The brass case becomes transparent for protons at about 55~MeV, approximately at the same energy as protons incident through the aperture start to penetrate the scintillator. RADMON registers protons in nine energy channels with threshold energies of 10 -- 40~MeV and electrons in five energy channels with threshold energies of 1.5 -- 12~MeV. The detailed analysis of the instrument response to electrons and protons is described in \citep{Oleynik-etal-2020}. The data rate can be adjusted by changing the polling frequency of the instrument. Nominally, science data is collected every 15 seconds and housekeeping data every 60 seconds. This gives a data rate of about 25.4 kBytes per hour, including the packet overhead.

Testing and ground calibrations of \gls{radmon} were performed using radioactive sources and a proton beam from the MGC-20 cyclotron at the \AA{bo} Akademi University, Turku, Finland. The maximum beam energy available in the cyclotron was about 17 MeV. The beam was scattered at about 60 degrees from a thin tantalum foil to lower the beam intensity and achieve a low-enough flux for the calibrations. The proton beam energy was step-wise decreased by adding absorbers between the foil and the detector. This setup allowed to successfully calibrate the instrument for the low-energy proton response, and Geant4 \citep{GEANT4-AGOSTINELLI, GEANT4-Allison} simulations were used to extend the proton response over the full energy range. The electron response was monitored utilizing beta particles from different radioactive decay sources.

Radiation tolerance of \gls{radmon} electronics has been tested in the \gls{radef} of the University of Jyvaskyla, Finland. The device was tested in a 50-MeV proton beam for total dose up to 10 krad, which it survived without observable degradation \citep{Peltonen2014}. As the device relies on a commercial version of the Xilinx Virtex-4 field-programmable gate array, we have implemented a triple-redundant memory with active scrubbing running parallel to the normal operations of the instrument \citep{Ilmanen2014}. The system was tested in \gls{radef} to be able to cope with a 50-MeV proton flux of $10^6$~cm$^{-2}$~s$^{-1}$, after which the rate of double bit errors became significant \citep{Peltonen2014}.

The instrument, being integrated into the satellite short stack, is also sensitive to electromagnetic interference. Especially the scintillator detector signal path is affected by the electromagnetic emission of other spacecraft subsystems. This has led to an increase of the noise levels in this signal and the inability to detect at the smallest signal levels, which has increased the threshold of the electron measurements from the nominal 0.7 MeV \citep{Peltonen2014} to 1.5 MeV achieved in space \citep{Oleynik-etal-2020}.

\subsection{Electrostatic Plasma Brake}

The plasma brake payload is based on the Coulomb drag principle, which is the driving phenomenon behind the Electric Solar Wind Sail (E-sail) invention \cite{angeo-25-755-2007, doi:10.2514/1.47537}. The brake itself consists of a 100 meter long tether; a storage reel; a vacuum qualified piezo motor and control electronics for tether deployment; a high voltage source; and four electron guns. Once the tether has been deployed, it can be charged with a voltage of either +1~kV or $-$1~kV, with respect to the surrounding ionospheric plasma. As the satellite moves through the plasma with its orbital speed, the electrostatic interaction between the tether and the plasma introduces a force opposite in direction to the satellite's velocity, thus slowly reducing the orbital speed.

\begin{figure}
    \centering
    \includegraphics[width=\textwidth]{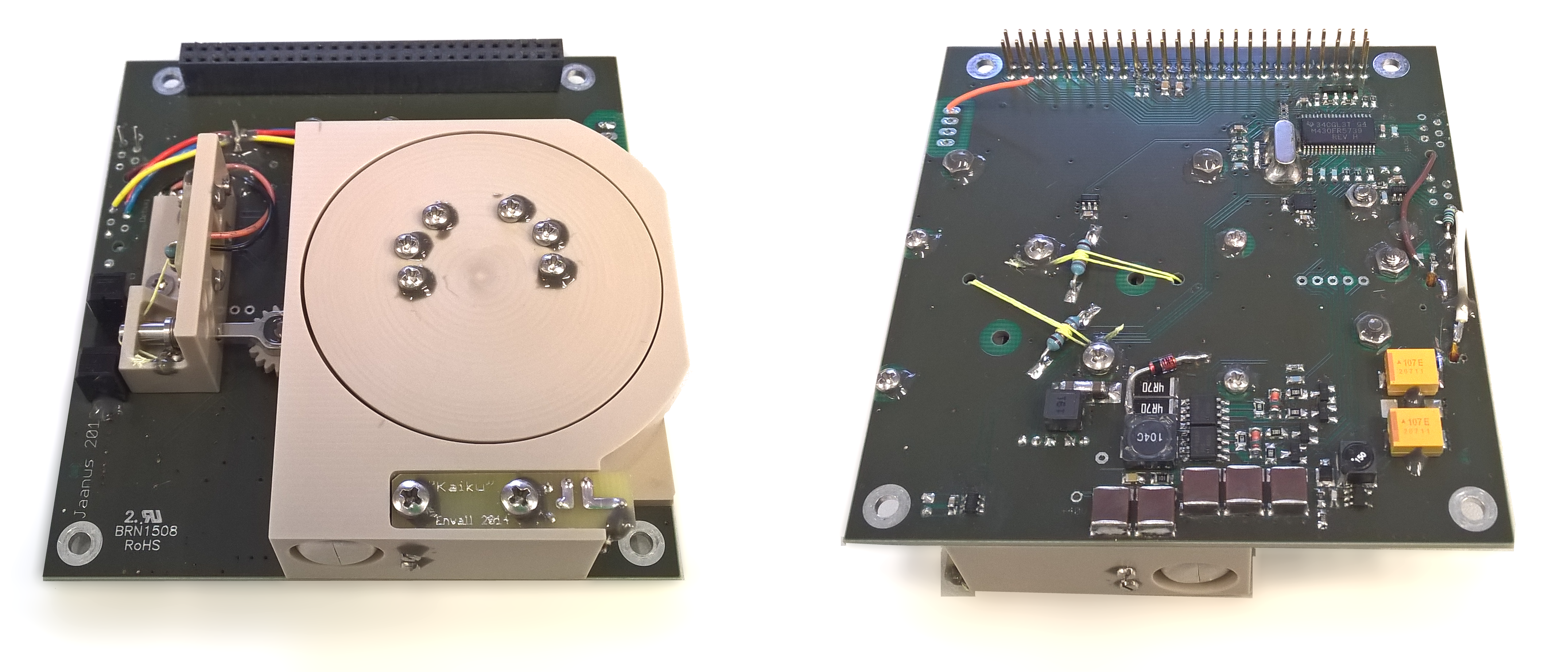}
    \caption{Tether {\jaan Reel} FM board from both sides. On the left, tether reel lock Kieku ready and locked. The black object left from Kieku is the optical feedback {\jaan Kyylä}. The electron guns are located satellite side panel X+, next to the payload which deploys from Z- end of the satellite.}
    \label{fig:tether_board}
\end{figure}

Fig.~\ref{fig:tether_board} shows the tether reel \gls{pcb}(left panel)  and the high voltage board (right panel). These two boards were stacked together to have dimensions of 5$\times$9$\times$10 cm. The spacecraft \gls{eps} supplied 3.3~V, 5~V, and 12~V to \gls{epb}. Power consumption of \gls{epb} depends on the operation mode: launch locks use 1.25~W each for about 20 sec (locks are not released at the same time), high voltage tether system consumes a few hundreds of mW depending on the ambient ionospheric plasma density, and the reeling system draws 2.3~W during the deployment. None of these tasks are executed simultaneously, and the tether is not deployed all at once. The data rate is low throughout the mission as only  the tether voltage and current are sampled with a frequency of 10~Hz. 

The tether itself is constructed of four aluminum filaments and it is based on the Heytether geometry \cite{SEPPANEN20113267}. The tether is deployed with the help of centrifugal force, the satellite must therefore be spinning around a suitable axis. Once the proper spin mode is reached, the tether reel motor is activated and the tether is slowly unreeled out to space. At the tip of the tether there is an aluminum tip mass, whose task is to assist in tether deployment by increasing the pull force experienced by the tether. On the bottom side of the reel there is a slip ring serving as the contact point for the high voltage source through two cantilever spring sliders being the only mechanically redundant subsystem of \gls{epb}. When a positive voltage is applied, one or more electron guns are activated in order to eject excess electrons and thus maintain the positive voltage, as the surrounding plasma attempts to neutralize it. In negative tether voltage mode, the tether gathers positive ions from the plasma and the conducting parts of the satellite surface collect the same flux of thermal electrons from the plasma to maintain current balance..

The plasma brake payload was tested prior to the system level tests. Vibration tests were carried out to qualify the mechanical components, \gls{pcb}s, and the reel motor, especially, as the motor was designed for laboratory use. It was noted that the high voltage sliders dug two dents to the slip ring that were able to stop reel rotation. Simple resistor-based launch locks were introduced to the bottom side of the reel \gls{pcb} to keep the sliders apart from the slip ring during the launch. The functionality of the payload was successfully tested in thermal-vacuum. Furthermore, specific to \gls{epb} payload, high voltage tests were made, and the tether outreeling was tested to determine the minimum centrifugal force required for the tether deployment.

\subsection{Aalto-1 Spectral Imager AaSI}

\begin{figure}[h!]
    \centering
    \includegraphics[width=9cm]{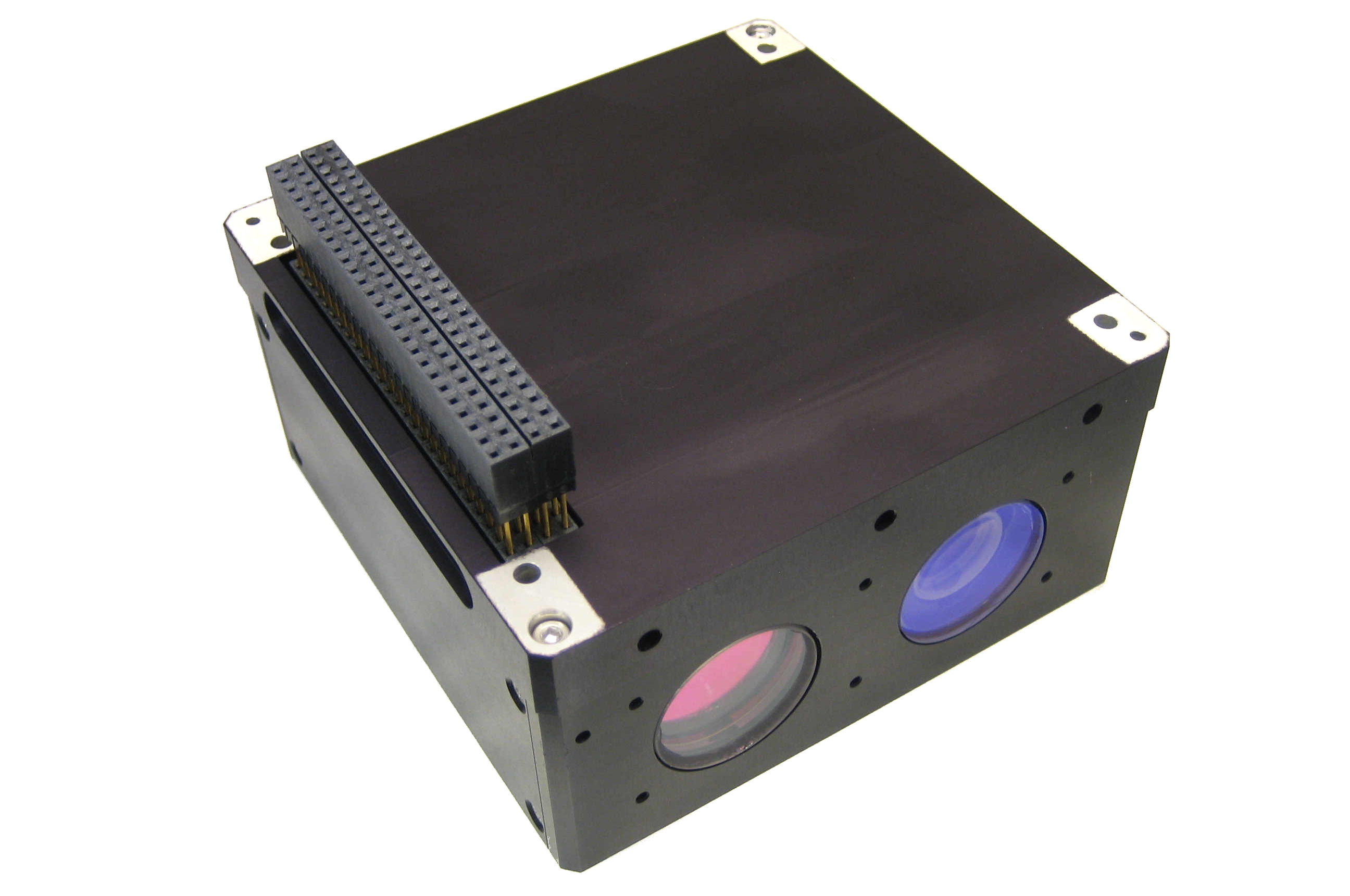}
    \caption{The Aalto-1 Spectral Imager AaSI. The size of the instrument is ca. 0.5 U and it is compatible with the PC104 interface. The instrument has two cameras: a visible spectrum RGB camera (left) and a spectral imager (right).}
    \label{fig:aasi1}
\end{figure}

The miniaturized spectral imager, AaSI, as shown in Fig.~\ref{fig:aasi1}, is the main payload of the \mbox{Aalto-1} nanosatellite. The imager is based on a tunable \gls{fpi}, which is used as an adjustable passband filter. This enables the imager to acquire images at freely selectable wavelengths. The operational range is 500--900~nm and the spectral resolution is 10--20~nm. In addition to the spectral imager, a visible (VIS) spectrum Red--Green--Blue (RGB) camera is included in the instrument \cite{NasilaMSc2013,Praks2018MiniatureSI, Praks18NanosatSpectral}.

\begin{table}[htb]
\begin{center}
\caption{Main parameters of AaSI.\label{AasiSpecs}}
\fbox{
\begin{tabular}{c|l}
\textbf{Wavelength range} & 500--900~nm  \\ \hline
\textbf{Spectral resolution} & 10--15~nm  \\ \hline
\textbf{Field of view} & 10$^\circ$ $\times$ 10$^\circ$ (SPE), 15$^\circ$ $\times$ 10$^\circ$ (VIS)    \\ \hline
\textbf{Spectral image size} & 512 $\times$ 512 pixels \\ \hline
\textbf{VIS image size} & 2048 $\times$ 1280 pixels \\ \hline
\textbf{Number of spectral bands} & 6, 25 or 75 \\ \hline
\textbf{Size} & 97 $\times$ 97 $\times$ 48 mm$^3$ \\ \hline
\textbf{Mass} & 600 g 
\end{tabular}
}
\end{center}
\end{table}

Table \ref{AasiSpecs} introduces the main parameters of \gls{aasi}.

\clearpage
\section{Platform}
\label{sec:platform}

The Aalto-1 platform subsystems include an \gls{eps} \cite{Hemmo13Master}, an \gls{adcs} \cite{Tikka14Attitude}, a GPS-based navigation system \cite{Leppinen13Master}, a \gls{uhf} \cite{Lankinen15Master,CanteroMSc2013} and S-band \cite{Jussila13SBand} radios for \gls{tt&c}, and a Linux-based \gls{obc} \cite{Razzaghi12Master,Javanainen16Master,SilvaMSc2017,Leppinen2019OBC}. The electronic subsystems are placed in two circuit board stacks, the Long Stack and the Short Stack, which are connected using a stack interface board. The electronics followed CubeSat electronics format, whereas the bus pin-layout followed PC-104 standard.  

The design philosophy of the CubeSat platform is a hybrid combination of subsystems developed in-house and commercial products. The satellite structure, solar panels, Sun sensors, \gls{tt&c} and \gls{obc} were fully designed in-house whereas the \gls{adcs} and the \gls{eps} were procured from commercial partners. The CubeSat structure, antenna and antenna deployment system were also developed in house. The in-house developed subsystems were fully designed, integrated and testing by student teams. The  \gls{pcb} designs were manufactured by commercial \gls{pcb} provider, whereas the component soldering and stuffing was performed in our facility. 
Special consideration was employed in the design of the critical subset of subsystems, consisting of \gls{eps}, \gls{obc} and \gls{uhf}. Redundant parts, fault detection and recovery procedures were added to increase their reliability and fault tolerance. 
The agile development approaches were followed in the design and verification of the satellite. The development process of the subsystems has been iterative, since the prototype of each subsystem was developed and qualified in quick iterations. The waterfall verification approach was followed in the Flatsat, \gls{eqm} and \gls{fm} integration \cite{Tikka2015}. The detailed design description of each platform subsystem is presented in the subsequent subsections. The in-orbit performance of major platform subsystems can be read from \cite{a1_inorbit_2020}.

\subsection{Electrical Power Subsystem}

\begin{figure}[h!]
    \centering
    \includegraphics[width=\columnwidth]{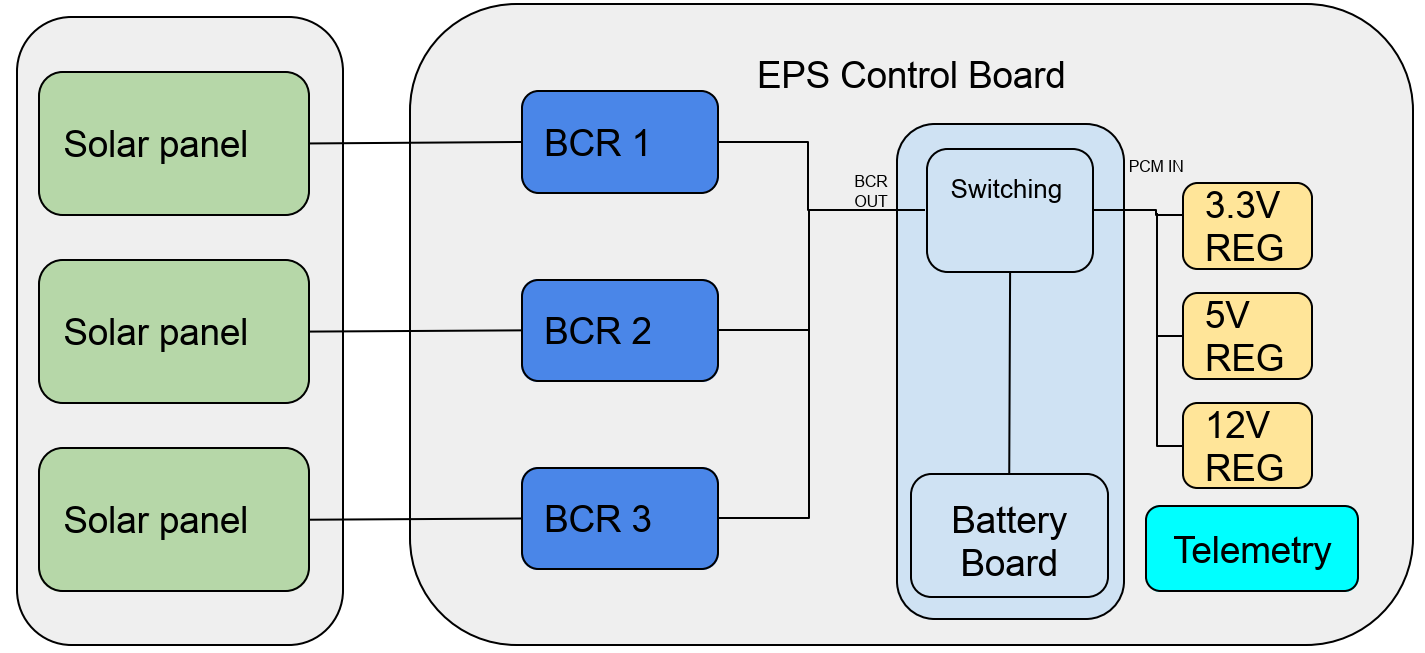}
    \caption{Block diagram of \gls{eps}}
    \label{fig:eps2}
\end{figure}

The \gls{eps} ensures the power generation, conditioning, storage and distribution to each subsystem and payload~\cite{Ali2013INNOVATIVEEP}. The \gls{eps} was procured from a commercial partner Clyde Space (Currently \r{A}AC-Clyde). The solar panels were designed in-house in order to accommodate the conductive surface requirements of \gls{epb}. A block diagram of Aalto-1 \gls{eps} is presented in Fig.~\ref{fig:eps2}

The incident solar radiation is converted to electrical power by the solar panels, developed at Aalto University \cite{FinnholmCubesat2013}. The Solar Panel design featured thermally conductive \gls{pcb} design. In-house made design provided freedom on sensor and payload location and satellite structure design.
The power from panels is transferred to the \gls{epscb} where \gls{bcr} convert the input voltage from the solar panels to battery charging voltage (6.2~V to 8.4~V). The \glspl{pcm} are responsible for regulating the voltages produced by solar panels and the battery unit. The dc-dc converters convert the voltage levels to the ones used by subsystems and distribute the power to the Satellite Bus (SB). The major subsystems have dedicated power lines, which are controlled by switches located on the battery board and accessible through stack connector. The operating voltages, standby and peak power consumption of platform avionics and payloads are provided in Table \ref{table:pbudget}.

\begin{table}[h!]
\caption{Aalto-1 Power budget. 
\label{table:pbudget}}
\centering

\begin{tabular}{|c|l|l|l|l|}
\hline
\multicolumn{1}{|l|}{System} & Details                                                          & \begin{tabular}[c]{@{}l@{}}Operating \\ voltage (V)\end{tabular} & \begin{tabular}[c]{@{}l@{}}Standby \\ Power (W)\end{tabular} & \begin{tabular}[c]{@{}l@{}}Peak \\ Power (W)\end{tabular} \\ \hline
\multirow{3}{*}{TT\&C}       & UHF                                                              & 12                                                               & 0.2                                                          & 1.55                                                      \\ \cline{2-5} 
                             & S-band                                                           & 3.3                                                              & 0                                                            & 3.5                                                       \\ \cline{2-5} 
                             & ADS                                                              & 12                                                               & 0                                                            & 7                                                         \\ \hline
\multirow{2}{*}{GPS}         & Active Antenna                                                   &                                                                  & 0.03                                                         & 0.03                                                      \\ \cline{2-5} 
                             & GPS Receiver                                                     &                                                                  & 0.015                                                        & 0.1                                                       \\ \hline
OBC                          &                                                                  &                                                                  & 0.25                                                         & 0.55                                                      \\ \hline
\multirow{2}{*}{ADCS}        & \begin{tabular}[c]{@{}l@{}}Coils and \\ Electronics\end{tabular} & 5                                                                & 0.5                                                          & 1.8                                                       \\ \cline{2-5} 
                             & Sun sensors                                                      & 5                                                                & 0                                                            & 0.06                                                      \\ \hline
\multirow{3}{*}{Payloads}    & RADMON                                                           & 12, 5                                                            & 0                                                            & 1                                                       \\ \cline{2-5} 
                             & EPB                                                              & 12, 5, 3.3                                                       & 2.3                                                          & 3                                                         \\ \cline{2-5} 
                             & AaSI                                                             & 12, 5                                                            & 0                                                            & 4                                                         \\ \hline
Total                        &                                                                  &                                                                  & 3.295                                                              & 22.59                                                           \\ \hline
\end{tabular}
\end{table}

The \gls{eps} has several safety features implemented for increased reliability of the platform. It monitors the \gls{i2c} bus lines for inactivity and erroneous behaviour (see Fig.~\ref{fig:obc_interfaces}), which if detected will cause a power cycle event of the whole platform. Battery power level is monitored as well and the low power mode is activated if depth of discharge is below the critical value. In this mode only the \gls{eps} is active, operating the battery charging circuits. Lastly, a timer feature, which starts a 30 minutes countdown after the first \gls{eps} power up, was set as a redundant antenna deployment trigger, in addition to the main dedicated countdown timers.

\subsection{Attitude determination \& control subsystem}

\begin{figure}[h!]
    \centering
    \includegraphics[width=0.7\columnwidth]{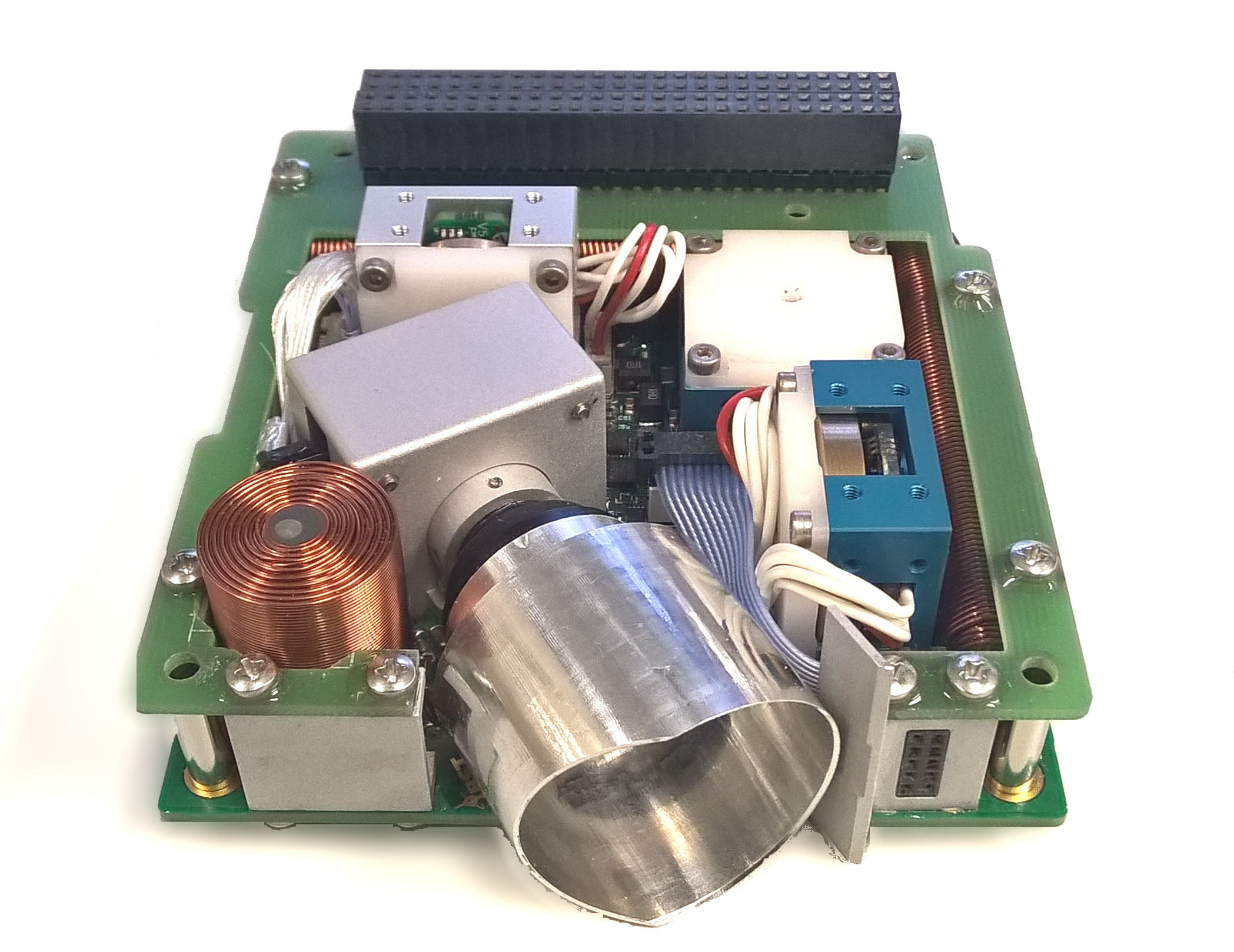}
    \caption{\gls{fm} of the \gls{adcs} iADCS100 with star tracker by Berlin Space Technologies GmbH and Hyperion Technologies.}
    \label{fig:adcs}
\end{figure}

The \gls{adcs} is the most critical subsystem to ensure the required pointing and spin modes for payloads. Aalto-1 \gls{adcs} (iADCS100), provided by commercial partners Berlin Space Technologies (BST) and Hyperion Technologies, consists of an integrated solution of attitude determination sensors and attitude control actuators. The attitude sensors include, gyroscopes, magnetometers and a star tracker. The Sun sensors were developed in-house by Aalto University and integrated to the solar panels\cite{TikkaJOSS}. The attitude actuators include magnetorquer rods and reaction wheels.Aalto-1 was the first satellite carrying iADCS100 attitude system and the Aalto students participated in the development. The \gls{fm} of Aalto-1 \gls{adcs} is shown in Fig.~\ref{fig:adcs}.

\subsection{Global  positioning system subsystem}

\begin{figure}[h!]
    \centering
    \includegraphics[width=\columnwidth]{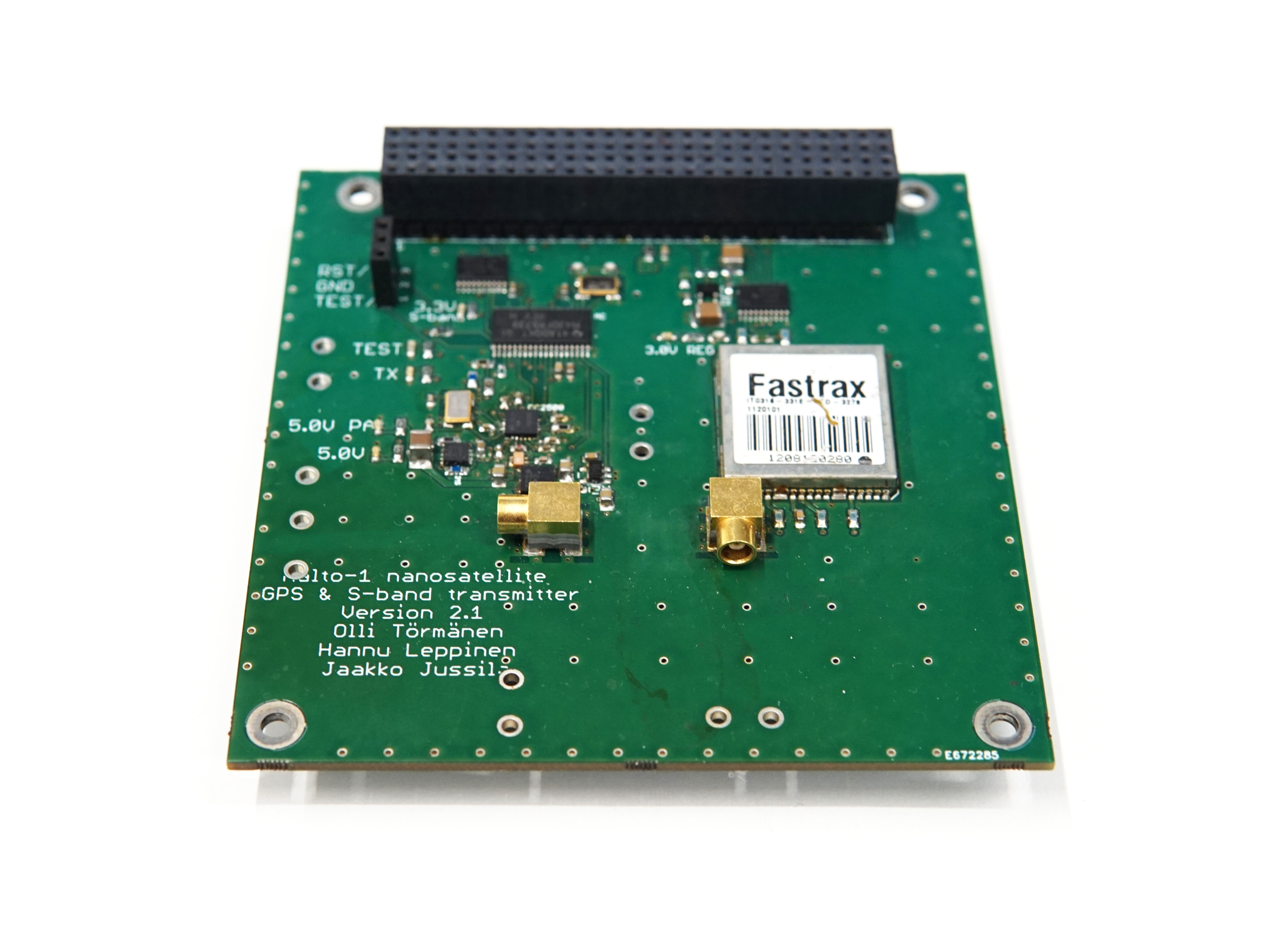}
    \caption{Aalto-1 S-band transmitter and GPS subsytem}
    \label{fig:gps}
\end{figure}

The GPS subsystem of Aalto-1 is shown in Fig.~\ref{fig:gps} which contains a Fastrax IT03 GPS receiver and an Adactus ADA-15S patch antenna. When operated, the GPS subsystem consumes approximately 160 mW of power \cite{Leppinen13Design, Leppinen13Master}. The main purpose of the subsystem has been to provide more accurate positioning than \gls{tle}-based solutions, for example, during plasma brake operations. The Fastrax receiver was selected as the manufacturer was willing to provide the receiver without the usual altitude and velocity restrictions \cite{Leppinen16Aalto1Nav}. In early 2010s, when a GPS subsystem was included in the Aalto-1 design, there were not many GNSS subsystems for nanosatellites available as commercial off-the-shelf products.

\subsection{Telemetry, tracking \& communication subsystem}

\begin{figure}[h!]
    \centering
    \includegraphics[width=\columnwidth]{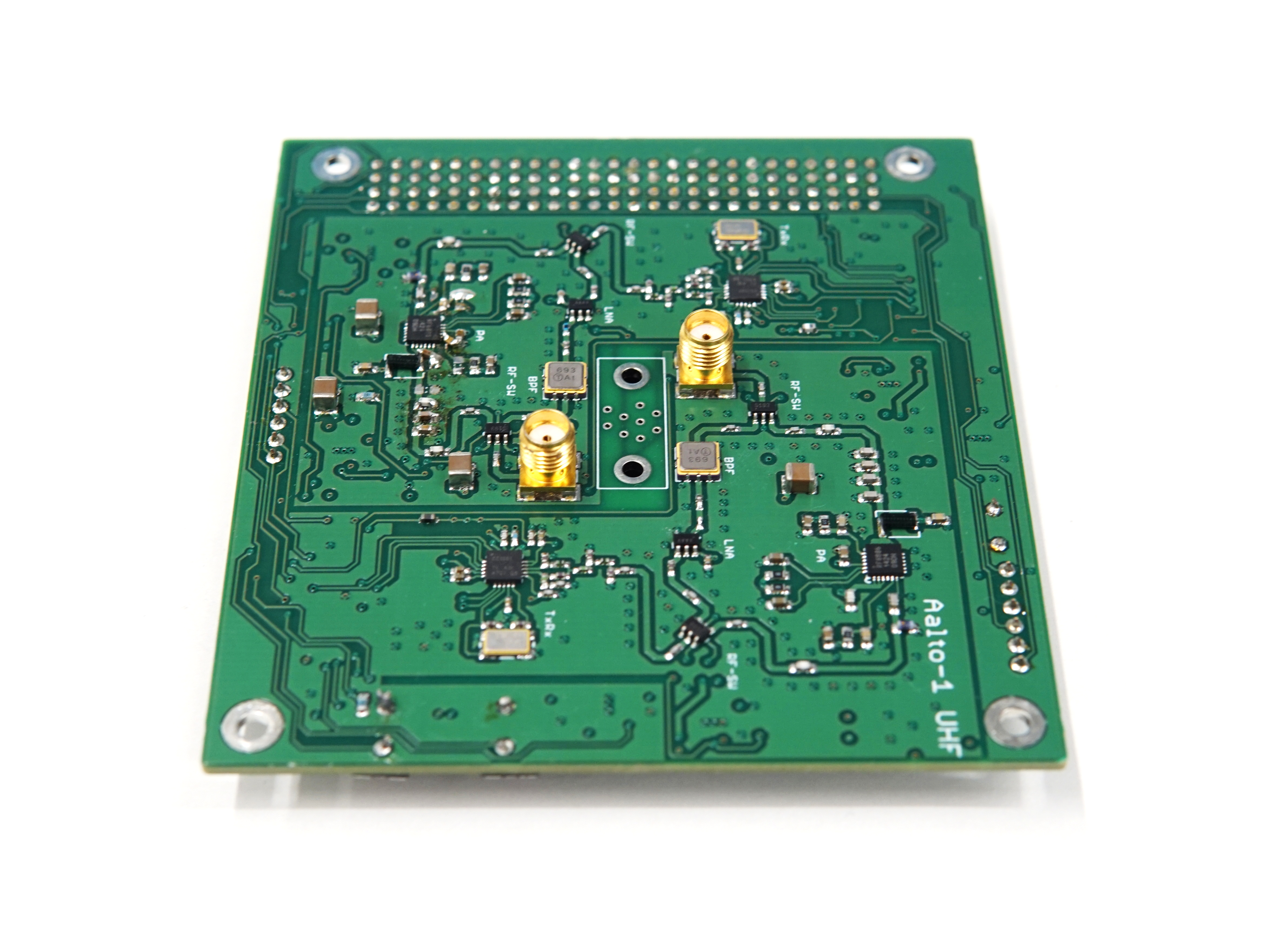}
    \caption{Aalto-1 UHF-band cold redundant transceiver}
    \label{fig:uhf}
\end{figure}

A \gls{uhf} transceiver was used as the primary radio on the Aalto-1 satellite. The \gls{uhf} radio supported transmission power of up to 1.2~W. The unit, as shown in Fig.~\ref{fig:uhf}, is fully redundant, equipped with two cold redundant TI CC1125-based transceivers and an MSP430 microcontroller (MSP430FR5729). It is capable of half-duplex bidirectional communication at 437.220 MHz. The \gls{uhf} communication system is equipped with two dipole \gls{uhf} antennas, each connected to one of the two redundant radios \cite{Lankinen15Master}. The \gls{obc} software and the arbiter can perform the switching from active to redundant radio.

\begin{figure}[h!]
    \centering
    \includegraphics[width=11cm]{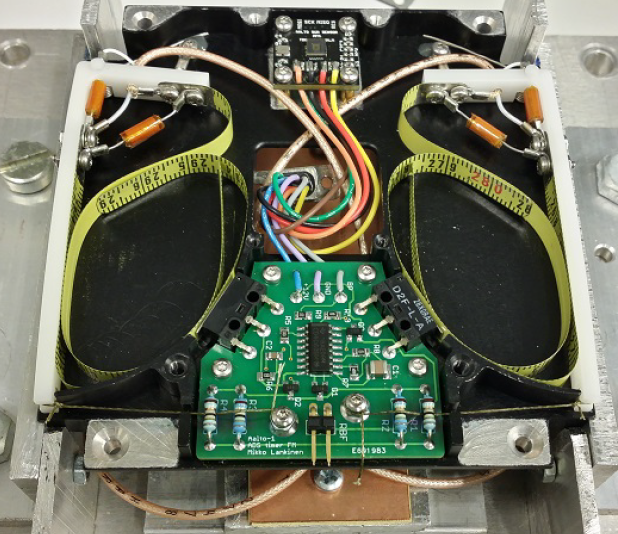}
    \caption{Aalto-1 Antenna deployment system and UHF-band antenna in stowed configuration}
    \label{fig:stowed}
\end{figure}

\begin{figure}[h!]
    \centering
    \includegraphics[width=\columnwidth]{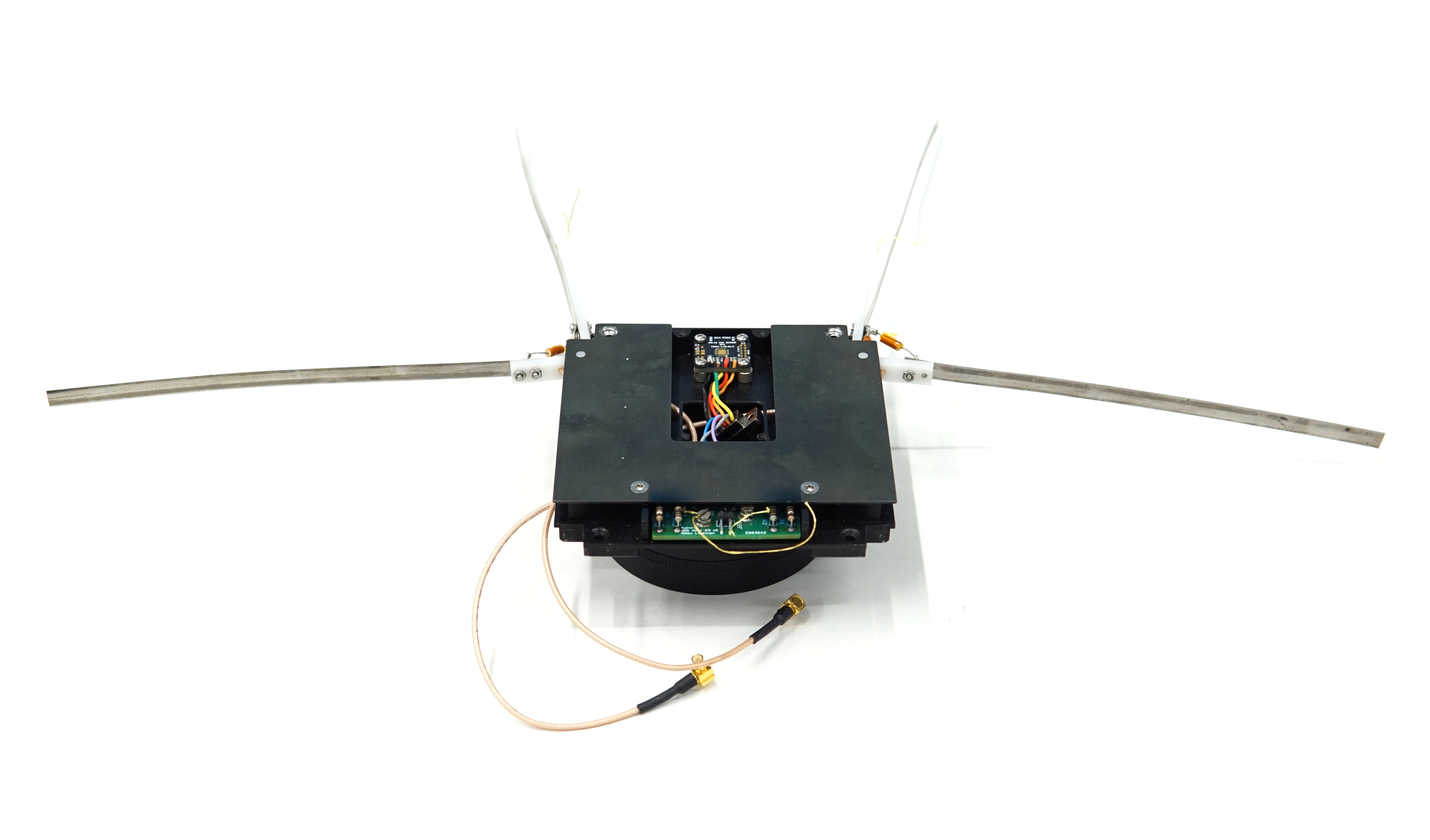}
    \caption{Aalto-1 UHF antenna in deployed configuration}
    \label{fig:deployed}
\end{figure}

A \gls{uhf} antenna deployment system, as shown in Fig.~\ref{fig:stowed}, consists of timer control board for antenna release and two L-shaped doors to keep the antennas stowed during launch. After the spacecraft is deployed, the antenna release mechanism burns the dyneema strings thereby deploying the antennas. Additionally, the redundant timer on the \gls{eps} can trigger the antenna deployment. The deployed antenna configuration is shown in Fig.~\ref{fig:deployed}.  
An automatic \gls{uhf} beacon is transmitted every two minutes by default. The \gls{uhf} beacon containing a static Morse code is transmitted every two minutes by default.

Along with the \gls{uhf} radio,  Aalto-1 has an S-Band transmitter used for high speed telemetry downlink. Because of regulations, the S-band transmission can be active only above the Aalto Ground Station. The S-band transmitter featuring a single transceiver (TI CC2500) and a microcontroller (MSP430FR5739) is shown in Fig.~\ref{fig:gps}. The communication frequency is 2.402 GHz with the design data rate of 500 kbps. The S-band communication system uses an in house designed single circular polarization patch antenna. It also forms a secondary downlink channel \cite{Jussila13SBand}.

\subsection{Onboard computer}
The Aalto-1 \gls{obc} consists of two cold-redundant 32-bit AT91RM9200, microcontrollers from Mircochip. The architecture hosts a 256-Mbit  \gls{sdram} volatile memory~ (AS4C16M16S), a parallel/NOR flash (S29JL064J), a dataflash (AT45DB642D), and a NAND flash (S34ML02G1). These different memories are used to store boot-loaders, kernel images and file systems. The architecture uses three different bus interfaces including \gls{i2c}, \gls{uart} and \gls{spi}. The \gls{uart}, \gls{spi} and USB are supported by the processor itself, while \gls{i2c} is handled by an an external controller (PCA9665).

\begin{figure}[h!]
    \centering
    \includegraphics[width=\columnwidth]{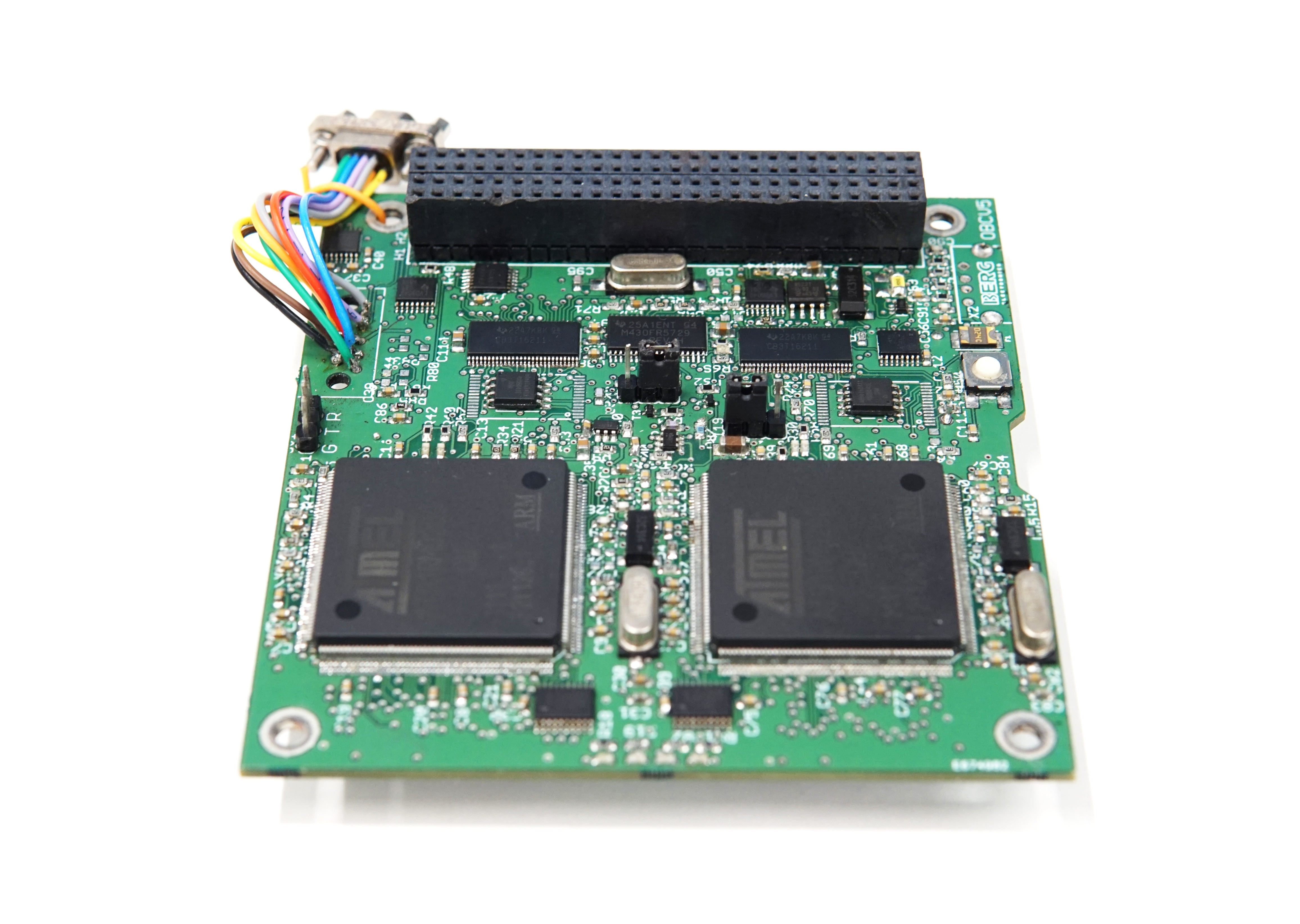}
    \caption{Aalto-1 On-Board Computer's flight spare}
    \label{fig:obc}
\end{figure}

The \gls{obc} consists of several components that can be classified as watchdogs, the most important one being the arbiter \cite{Javanainen16Master}. An MSP430-based arbiter selects which of the two processors is powered therefore preventing mission failure due to hard failure of one of the \glspl{obc}.
In the arbitration logic, a full reboot is required to switch from the active to redundant \gls{obc}. The switching procedure was set to execute when the arbiter powers up and it does not receive a heartbeat signal of the active \gls{obc}. A further read on system description, arbitration logic, \gls{fmea} and error handling procedures, can be found in \cite{Razzaghi12Master}. A detailed block diagram of the \gls{obc} representing the data interfaces with payloads and platform subsystems is shown in Fig.~\ref{fig:obc_interfaces} whereas the flight spare model of the \gls{obc} is shown in Fig.~\ref{fig:obc}.

The \gls{obc} runs the Linux operating system and bash for scripting different command sequences. At the time of its selection in 2010, Linux was not a common choice for satellite \gls{obc}s, but has since become popular in small spacecraft \cite{Leppinen17Linux}, \cite{Leppinen2019OBC}.
Software of the \gls{obc}, due to high complexity of the Linux operating system, has been thoroughly analysed and additionally strengthened against various identified failure scenarios\cite{Javanainen16Master}. The version control in software development approach was followed in the software design with  regular commits to the Github repository.

\subsection{Software}

Several on-board data handling tools have been used in existing CubeSat designs \cite{ali2012uml}. The Aalto-1 on-board data handling and flight software is built around applications running on Linux which was quite a new choice for nanosatellites at the design selection stage. The applications utilise certain libraries to communicate with satellite subsystems and the satellite internal data bus.

A number of libraries were developed for several subsystems, the most prominent being libarbiter for arbiter, libeps for communication with \gls{eps}, libicp for communications with subsystems on \gls{i2c} and libradio and libsband for radio communication. The detailed description on design choices and lessons learned on Aalto-1 software design approach can be followed in \cite{Leppinen2019OBC}.

Linux is a feature full operating system with mature, stable and time-proven core code base. This simplified the development of flight logic and utilities. Well known Linux library ecosystem and APIs assured a proper separation of concerns. 

Nonetheless, Linux is a complex software which necessitated a thorough analysis to ensure reliability on the \gls{obc}\cite{Javanainen16Master}. During the system's boot procedure there is no possibility for intervention and thus needed to be made fault tolerant by adding an emergency boot procedure with reduced functionality and less dependencies. A memory storage was divided into sections with primary and recovery file systems. Unsorted Block Image File System (UBIFS) have been used and it supports wear leveling and due to its use of journals is power loss tolerant. On the overall system level, a number of software and hardware watchdog timers are used in conjunction with the arbiter heartbeat output. Bus and radio communication libraries are strengthened with appropriate checksum and implementing non-blocking procedures.

\subsection{Thermo mechanical subsystem}

There are two long and a short standard PC-104 stack to route power and data signals among platforms and payloads . The long stack, required by few subsystems, is 2U long whereas the short stack is 1U long. 
As evident from Fig.\ref{fig:a1internals}, the orientation of subsystems on one unit is different than those on the other two units, therefore a stack interface board was used. 
The in-house built structure is compatible with standard dimensions and provides mechanical interface to internal subsystems, solar panels and antenna deployment. The breakdown of total spacecraft mass is provided in Table \ref{table:mbudget}

\begin{table}[h!]
\caption{Aalto-1 mass budget
\label{table:mbudget}}
\centering
\begin{tabular}{|l|l|l|l|l|}
\hline
System                                       & Details               & Quantity & Unit mass (g) & Total mass (g) \\ \hline
Structure                                    & 3U and harness                    & 1        & 1180          & 1180           \\ \hline
\multicolumn{1}{|c|}{\multirow{4}{*}{TT\&C}} & UHF antenna           & 4        & 1             & 4              \\ \cline{2-5} 
\multicolumn{1}{|c|}{}                       & UHF transceiver       & 1        & 90            & 90             \\ \cline{2-5} 
\multicolumn{1}{|c|}{}                       & S-band \& GPS board   & 1        & 75            & 75             \\ \cline{2-5} 
\multicolumn{1}{|c|}{}                       & S-band antenna        & 1        & 50            & 50             \\ \hline
\multicolumn{1}{|c|}{\multirow{3}{*}{EPS}}   & Solar panels          & 4        & 130           & 520            \\ \cline{2-5} 
\multicolumn{1}{|c|}{}                       & Control board         & 1        & 83            & 83             \\ \cline{2-5} 
\multicolumn{1}{|c|}{}                       & Batteries             & 1        & 258           & 258            \\ \hline
OBC                                          &                       & 1        & 75            & 75             \\ \hline
\multirow{2}{*}{ADCS}                        & Coils and Electronics & 1        & 360           & 360            \\ \cline{2-5} 
                                             & Sun sensors           & 6        & 10            & 60             \\ \hline
\multirow{3}{*}{Payloads}                    & RADMON                & 1        & 360           & 360            \\ \cline{2-5} 
                                             & EPB                   & 1        & 300           & 300            \\ \cline{2-5} 
                                             & AaSI                  & 1        & 600           & 600            \\ \hline
Total                                        &                       &         & 3572               & 4015          \\ \hline
\end{tabular}
\end{table}

The spacecraft used passive thermal control system \cite{9160911}. The structure rails were anodized black since it provides optimum emissivity/absorptivity ratio. The electrically conductive surfaces were masked before anodization and later chromate coated. The unused areas of solar cell \gls{pcb}s were gold plated. In order to increase the thermal conductivity from solar cell to the structure, indium foil washers were placed in the screw joints. For better thermal conductivity, many grounding vias were also placed in the solar cell footprints. The telemetry data of Aalto 1 reveals that the equilibrium temperature is well maintained inside the spacecraft.

\section{Satellite integration \& testing}
\label{sec:integration}
The model philosophy of the project followed a Flatsat, \gls{eqm} and \gls{fm} approach. This approach was selected mainly due to the fact that all subsystems and payloads were new development items and early verification of them was seen as highly beneficial. Additionally, lessons learned from other CubeSat projects in other universities often highlighted the importance of leaving significant amount of time for the integration and testing campaign on the system level prior to a launch.

\begin{figure}[h!]
    \centering
    \includegraphics[width=\columnwidth]{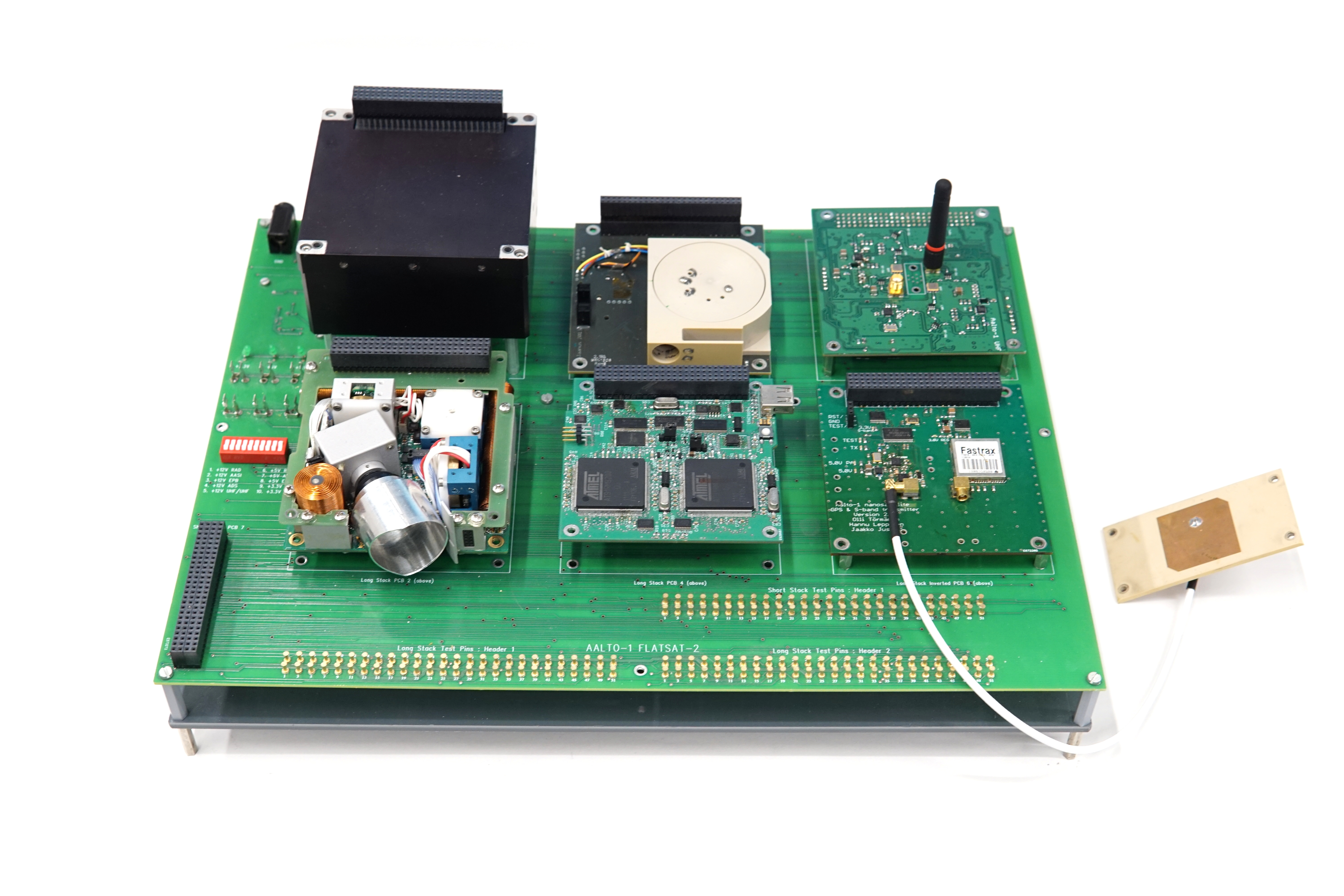}
    \caption{Aalto-1 Flatsat model integrating (starting from top left continuing clockwise)  \gls{aasi}, \gls{epb}, \gls{uhf}, S-band with patch antenna and \gls{gps}, \gls{obc} and commercial \gls{adcs}}
    \label{fig:flatsat}
\end{figure}

Most of the testing performed prior to system level integration was done on subsystem level by each development team. Usually, development kits and other test equipment was utilized rather than other satellite subsystems as their development was performed concurrently by separate teams.The first full interface tests were performed on the Flatsat model which is shown in Fig.~\ref{fig:flatsat}.  A number of interface mismatches were identified and troubleshooted at this stage. While being rather typical for any project with many concurrent developments, earlier testing of the system as a whole, if possible, would probably have saved required redesign and manufacturing effort at the \gls{eqm} level. As most of the satellite subsystems were in-house developed, making small modifications was however relatively easy at this point of the project.

A full environmental qualification test campaign was performed with the satellite \gls{eqm}. No major issues were found during these tests, which raised the confidence level on the system level design. However, not all satellite functionalities had been implemented at this point and thus were not fully reference tested prior and during the environmental testing campaign. This left some uncertainties to be fully verified later at the \gls{fm} test campaign. It also highlighted the importance of thorough reference testing and reporting prior to environmental tests.

The satellite \gls{fm} was built soon after the \gls{eqm} environmental test campaign, including some necessary minor modifications. Testing with the \gls{eqm} also continued throughout the \gls{fm} campaign and allowed simpler software development and testing on the system level, as well as tests that could have caused unnecessary stress to the \gls{fm}. Such tests were, for example, long duration durability testing, outdoor long-range testing and magnetic testing. Some issues were still found using the \gls{eqm} and it was possible to implement necessary fixes to the \gls{fm}. One of such issues was related to a component in the Telemetry/Telecommand (TM/TC) radio and may not have been noticed without the long duration durability testing, and could possibly have caused mission failure soon after the launch.

A full acceptance test campaign was performed with the satellite \gls{fm}. The \gls{fm} testing consisted of pre-built test scripts to command the subsystems and receive respective telemetries. No major issues were found during these tests, as was expected thanks to the successful \gls{eqm} tests.  Due to the late readiness of the third party provided \gls{adcs} and flight software, it was not possible to perform a thorough enough functional or performance test campaign for it. Testing of the \gls{adcs} algorithms was planned to be performed using a hardware-in-loop approach, which did not work as expected through the satellite main communication bus due to communication delays. Rather, access to the \gls{adcs} internal sensor and actuator bus would have been required, but it was not possible at that point. This highlighted the importance of early delivery of third-party systems with final and fully tested flight software and should be considered a high risk regarding any new developments by a third party. 

The Aalto-1 launch campaign started after the assembly, integration and verification stage. A lot of issues were addressed even when the satellite was in the launch pod. As an example, the batteries had become empty and it was a trouble charging them because no such interface was provided on the access port. The batteries in \gls{fm} were charged with a solar lamp transported to the launch pod.

\begin{figure}[h!]
    \centering
    \includegraphics[width=\columnwidth]{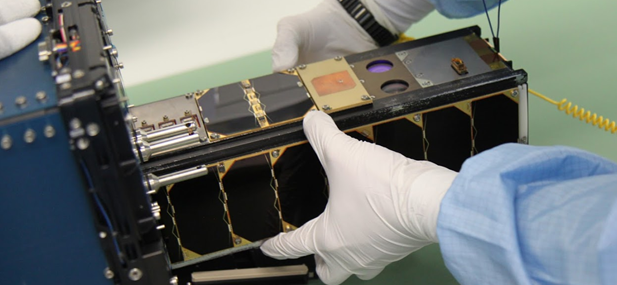}
    \caption{Aalto-1 integration with the deployer}
    \label{fig:integration}
\end{figure}

A photograph of the integration of the \gls{fm} into the commercial orbital deployer is shown in Fig.~\ref{fig:integration}.

In the end, the selected model philosophy proved to be a very suitable approach for the project that had significant amount of new development items. Like in many other CubeSat projects, schedule issues were encountered to perform system level testing in the most ideal and thorough way possible. Emphasis on system level testing from the very beginning of development can solve some of such issues encountered at later stages of the project. A most-viable-product approach, used typically in software engineering, has been followed and determined beneficial in projects after Aalto-1, where the highest importance functions of the satellite are implemented and tested as early as possible on the system level, and later incremented with additional features in order of priority towards the full satellite integration and testing. Such an approach however requires agile in-house development and close cooperation with third parties. Ultimately, the most suitable development approach for any CubeSat project is highly dependant on many aspects, such as the available resources, experience, the number of development items and the usage of in-house or third-party systems. The development approach should be carefully planned only after such aspects have been identified.

\section{Ground segment \& services}
\label{sec:ground}

The ground segment originally used an Icom IC-910H radio transceiver and a relay based pre-amplifier that was designed for voice communication. Reception was implemented using an RTL-SDR.
Five months after launch, the setup was updated with a newly developed solid state switched pre-amplifier due to problems with the relay-switched pre-amplifier.

In 2018, the transceiver was changed to a USRP B200 \gls{sdr}. To ease operation of multiple missions from the ground station, the digitized radio signal is distributed to multiple programs through a shared memory buffer. With the OpenWebRX software, the spectrum between \SI{431}{\mega\hertz} and \SI{439}{\mega\hertz} can be monitored using a web browser.

Brushed motors in the antenna rotator caused strong, broadband interference close to the antenna while rotating during a satellite pass. The issue was reduced with upgraded rotators and an upgraded controller.

\begin{figure}[h!]
    \centering
    \includegraphics[width=\columnwidth]{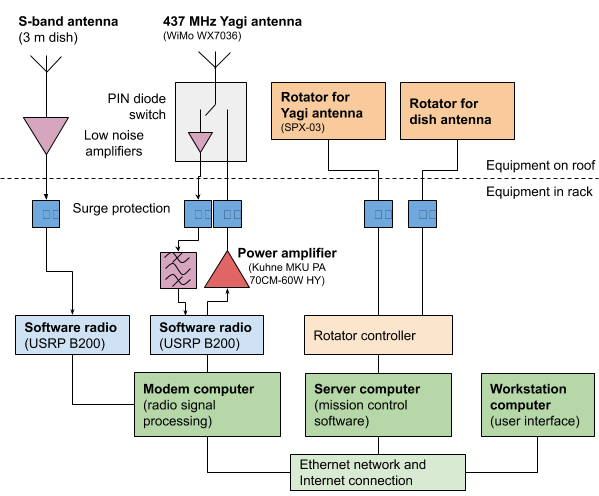}
    \caption{A block diagram of Ground station.}
    \label{fig:a1_gs}
\end{figure}

Block diagram of the relevant parts of the currently operational ground station is presented in Fig.\ref{fig:a1_gs}.

The ground segment is controlled using the \gls{mcc} software developed by the Aalto-1 team. The back-end is based on a PostgreSQL database that stores every received packet with a timestamp. Furthermore, the housekeeping system stores every housekeeping value separately with timestamp which has proven cumbersome and ineffective due to storage of large set of values in the database. For upcoming missions, we plan to use a database structure that uses one table per subsystem and one row includes an entire housekeeping package of that subsystem. This should reduce query time substantially as not every value has to be queried independently.

We have developed a \gls{gui} for the \gls{mcc} based on Qt. Qt was chosen since the application programming interface (API) does not change very quick which ensures long compatibility in the future. The \gls{gui} shows information about the current position of the satellite, satellite passes, received and transmitted packets and housekeeping most recent data. In addition, the history of a single housekeeping value can be plotted.

\section{Conclusion}
\label{sec:discussion}

The Aalto-1 projects key scientific and technological and educational objectives were achieved. The platform and payloads were successfully designed, developed and integrated with many student teams getting hands-on learning. The integration, testing, verification and launch activities were successfully accomplished. The subsystems and payloads demonstrated partial mission success with many lessons learned which have been briefed in an accompanying paper.

This project started a new era of space activities in Finland. A number of new space start-ups were founded as an outcome of this project. The (former) Aalto satellites group members have started and joined a number of new missions, such as ICEYE SAR satellite constellation, Aalto-3, Reaktor Hello World, FORESAIL~\cite{Palmroth_2019}, and Comet Interceptor~\cite{Snodgrass2019TheWait}. The Aalto-1 design has been been beneficial in the space technology curriculum and a source of inspiration for new students in the space technology lab. 

\section*{Acknowledgements} 
The RADMON team thanks P.-O.\ Eriksson and S.\ Johansson at the Accelerator Laboratory, \AA{}bo Akademi University, for operating the cyclotron. Testing work at the University of Jyvaskyla has been supported by the Academy of Finland under the Finnish Centre of Excellence Programms 2006-2011 and 2012-2017 (Project No:s 213503 and 2513553, Nuclear and Accelerator Based Physics), and by the European Space Agency (ESA/ESTEC Contract 18197/04/NL/CP).

Aalto University and its Multidisciplinary Institute of Digitalisation and Energy are thanked for Aalto-1 project funding, as are Aalto University, Nokia, SSF, the University of Turku and RUAG Space for supporting the launch of Aalto-1.

\bibliographystyle{model1-num-names}
\bibliography{ms}

\end{document}